\documentclass[10pt,twocolumn,showpacs,preprintnumbers,amsmath,amssymb,aps,prb,
superscriptaddress,floatfix,footinbib]{revtex4-1}

\usepackage{graphicx}
\usepackage{amsmath}
\usepackage{amssymb}
\usepackage{hyperref}

\usepackage[active]{srcltx}
\usepackage{lipsum}
\usepackage{color}
\def\tr{\textcolor{red}}
\newif\iftransport
\transporttrue
\newif\ifMPG
\MPGfalse
\newif\iftodo

\usepackage[usenames,dvipsnames,svgnames,table]{xcolor}

\begin{document}


\title{Magneto-optic and transverse transport properties 
       of non-collinear antiferromagnets} 
\author{Sebastian Wimmer} 
\email{sebastian.wimmer@cup.uni-muenchen.de}
\affiliation{%
  Department  Chemie,  Physikalische  Chemie,  Universit\"at  M\"unchen,
  Butenandtstr. 5-13, 81377  M\"unchen, Germany\\}
\author{Sergiy Mankovsky}
\affiliation{%
  Department  Chemie,  Physikalische  Chemie,  Universit\"at  M\"unchen,
  Butenandtstr.  5-13, 81377 M\"unchen, Germany\\}
\author{J\'an Min\'ar}
\affiliation{%
 New Technologies Research Centre, University of West Bohemia, Univerzitni 8,
30614 Pilsen, Czech Republic\\}
\author{Alexander N. Yaresko}
\affiliation{%
  Max-Planck-Institute for Solid State Research, Heisenbergstrasse 1, 70569 Stuttgart, Germany\\}
\author{Hubert Ebert}
\affiliation{%
  Department  Chemie,  Physikalische  Chemie,  Universit\"at  M\"unchen,
  Butenandtstr. 5-13, 81377 M\"unchen, Germany\\}


\begin{abstract}
Previous studies on the anomalous Hall effect in coplanar non-collinear 
antiferromagnets are revisited and extended to magneto-optic properties, namely 
magneto-optic Kerr effect (MOKE) and X-ray magnetic dichroism (XMCD). Starting 
from group-theoretical considerations the shape of the frequency-dependent conductivity 
tensor for various actual and hypothetical spin configurations in cubic 
and hexagonal Mn$_3X$ compounds is determined. Calculated MOKE and X-ray 
dichroism spectra are used to confirm these findings and to give estimates of 
the size of the effects. For Mn$_3$IrPt and Mn$_3$PtRh alloys the concentration 
dependence of the anomalous and spin Hall conductivity is studied in addition.
\end{abstract}

\maketitle

\section{Introduction}

The anomalous Hall effect (AHE) in magnetically ordered materials is usually 
considered to scale with the corresponding 
magnetization.\cite{PR53,YKM+04,ZYNW06,NSO+10} The same applies to the 
magneto-optical Kerr effect\cite{Ker78,Arg55} (MOKE) that is used in particular 
to monitor the magnetization dynamics.\cite{BMDB96,WMHB07} While the linear 
relationship of the AHE and the MOKE with the magnetization appears plausible, 
there seems to be no strict formal justification given for this in the 
literature apart from numerical studies based on \emph{ab initio} 
calculations.\cite{TFYK05} Nevertheless, a sum rule has been derived that 
relates the integrated off-diagonal optical conductivity to the orbital magnetic 
moment of the material.\cite{HH61,Smi76} The magnetic circular dichroism in 
X-ray absorption (XMCD) is very closely related to the MOKE.\cite{Ebe96} To make 
full use of this local magnetic probe several authors have derived the so-called 
XMCD sum rules \cite{ES75,TCSL92,CTAW93} that allow for example to deduce from 
the integrated L$_{2,3}$-spectra of 3d-transition metals their spin and orbital 
magnetic moments. In line with the sum rules an angular dependency according to 
$\cos(\hat m \cdot \hat q)$ is assumed, where $\hat m$ and $\hat q$ are the 
orientation of the local moment probed by XMCD and that of the incident X-ray 
beam. This simple relation implies in particular that in spin-compensated 
antiferromagnetic systems the XMCD should vanish. For such systems information 
on the local magnetic moment can nevertheless be obtained by exploiting the 
linear magnetic X-ray dichroism (XMLD)
for which two spectra with linear polarization parallel and perpendicular to the 
local magnetization are recorded.\cite{CKTA93}

In contrast to this situation, Chen \emph{et al.}\cite{CNM14} pointed out
that perfect spin-compensation in an antiferromagnet is not a sufficient
criterion for the AHE to be absent. In fact these authors  give symmetry
arguments that the AHE can indeed   occur for example in spin-compensated
non-collinear antiferromagnets as  Mn$_3$Ir that is commonly employed in
spin-valve devices.  Numerical work by  Chen \emph{et al.} for  Mn$_3$Ir
gives in fact a rather large anomalous Hall conductivity, comparable in size
to the values for Fe, Co, and Ni. K\"ubler and Felser\cite{KF14a} numerically
confirmed the results of Chen \emph{et al.} and investigated in addition the
non-collinear antiferromagnetic hexagonal compounds Mn$_3$Ge and Mn$_3$Sn
considering various coplanar and non-coplanar triangular magnetic
configurations. Also in this case several spin-compensated configurations
were identified that were predicted to exhibit an AHE. 
These predictions recently could indeed be experimentally verified in
Mn$_3$Sn\cite{NKH15} and Mn$_3$Ge\cite{KTN16,NFS+16}.

Here one should stress that the occurrence of off-diagonal anti-symmetric 
conductivity tensor elements has been unambiguously determined much earlier. In 
particular Kleiner\cite{Kle66} gave the space-time symmetry-restricted tensor 
forms for the electrical conductivity in all magnetic solids based on the 
transformation properties of the corresponding Kubo formula under the symmetry 
operations of the relevant magnetic group. These results in particular do not 
rest on the assumption of collinear magnetic order or a finite magnetization. 
Moreover, they apply to the frequency-dependent or optical conductivity just as 
well and therefore imply the presence of MOKE as well as XMCD signals 
concomitant with the AHC. The former has been studied in chiral cuprates based on 
tight-binding model calculations of the Berry curvature\cite{OM13} and, using 
first-principles methods, in cubic Mn$_3$Ir-type compounds\cite{FGZ+15a}. The 
symmetry criteria for the presence of the spin Hall effect (SHE) have recently 
been given by the present authors\cite{SKWE15a}, extending the approach by 
Kleiner\cite{Kle66} to more complex situations. The SHE in non-collinear 
antiferromagnets is currently the subject of intensive theoretical 
efforts\cite{Gom15,ZSY+17,ZZFY17a,ZZS+17} and has been observed in experiments 
on Mn$_3$Ir\cite{ZHY+16a} via its contribution to the so-called spin-orbit 
torque. Due to the identical transformation properties of electronic charge and 
heat current operators\cite{Kle66,SKWE15a} the occurrence of AHE and SHE imply 
the existence of corresponding thermally-induced effects, the anomalous and spin 
Nernst effects. These have been studied in  hexagonal Mn$_3X$ ($X =$
  Sn, Ge, Ga) using the Berry 
curvature approach and a Mott-like formula.\cite{GW17} The anomalous Nernst 
effect has recently been observed experimentally in Mn$_3$Sn.\cite{ITK+17} Of 
particular current interest in the thriving field of non-collinear magnetism is 
the occurrence of chirality-induced or so-called \emph{topological} effects 
arising from the emergent electromagnetic field created by a non-trivial real- 
or reciprocal-space texture.\cite{NT13} This will be the subject of another 
publication.\cite{WME18}

In the following we present a theoretical study on the transverse transport
and optical properties of non-collinear coplanar antiferromagnets. The next
section (\ref{sec:theo}) gives a brief overview on the theoretical framework
and methods applied. The major part of this contribution is devoted to results
for the cubic prototype system Mn$_3$Ir in section~\ref{sec:Mn3Ir} and a number 
of possible magnetic configurations of hexagonal Mn$_3$Ge in section~\ref{sec:Mn3Ge}.

\section{Theoretical framework \label{sec:theo}}





\subsection{Magnetic symmetry and shape of the conductivity tensor}

For the specific case of Mn$_3$Ir  the occurrence of a non-vanishing AHE was 
predicted by Chen \emph{et al.}\cite{CNM14} by explicitly considering 
Kagome-type sub-lattices occupied by Mn-atoms with triangular antiferromagnetic 
order (see below) in combination with a suitable model Hamiltonian. A more 
general scheme to search for a finite AHE in spin-compensated systems is to use 
Kleiner's tables that give among others the shape of the frequency dependent 
conductivity tensor {\boldmath$\sigma$}$(\omega)$ for any material.\cite{Kle66} 
These tables were constructed by starting from Kubo's linear response formalism 
and making use of the behavior of the current density operator under all 
symmetry operations of the relevant magnetic space group. It turns out that only 
the magnetic Laue group of a material has to be known to fix unambiguously the 
shape of {\boldmath$\sigma$}$(\omega)$. Unfortunately, Kleiner\cite{Kle66} used 
an old definition for the Laue group that is obtained by removing the inversion 
from all improper symmetry operations of the magnetic point group leading to the 
magnetic Laue Group $32'$ in the case of Mn$_3$Ir.\cite{Kle66} With the 
(magnetic) Laue group defined\cite{Bor12} to be the (magnetic) point group 
artificially extended by the inversion one is led to the magnetic Laue group 
$\overline{3}m'$ instead. Kleiner's tables have been updated recently by Seemann 
\emph{et al.}\cite{SKWE15a} to account in particular for the revised definition 
of the magnetic Laue group.

Having determined the magnetic Laue group of a solid --for this purpose the program 
FINDSYM\cite{ISOTROPY,SH05} was used here-- the specific shape of its 
conductivity tensor  {\boldmath$\sigma$}$(\omega)$ can be read from these 
tables. The presence of an anti-symmetric part for the off-diagonal tensor 
elements in particular indicates the simultaneous occurrence of the AHE, the 
MOKE as well as the XMCD (see below).

\subsection{First-principles calculations of the anomalous Hall effect, the magneto-optical Kerr effect and the X-ray magnetic dichroism}

The qualitative investigations on the occurrence of the AHE, the MOKE and XMCD 
of various materials presented below, had been complemented by corresponding 
numerical work. The underlying electronic structure calculations have been done 
within the framework of relativistic spin density functional theory with the 
corresponding Dirac Hamiltonian given by:\cite{MV79}
%
%
\begin{equation}
\label{eq:Dirac}
{\cal H}_{\rm D} = c \mbox{\boldmath{$\alpha$}}  \cdot {\vec p}
 + \beta mc^2 +V(\vec{r}) +\beta \Sigma_z B(\vec{r}) 
 \; .
\end{equation}
%
Here all quantities have their usual meaning \cite{Ros61}, with the 
spin-averaged and spin-dependent exchange correlation contributions  $\bar 
V_{\rm xc}(\vec{r})$ and $B_{\rm eff}(\vec{r})$, respectively, to the effective 
potential $V(\vec{r})$ set up using the parametrization of Vosko {\em et al.} 
\cite{VWN80}. To deal with the resulting four-component Dirac equation the 
spin-polarized relativistic (SPR) formulations of the Korringa-Kohn-Rostoker 
(KKR) \cite{EBKM16,SPRKKR} and linear-muffin-tin-orbital 
(LMTO)\cite{Ebe88,PY-LMTO,AHY04} methods have been used. With the electronic 
Green Function   $G^+(\vec{r},\vec r\,',E)$ supplied by the SPR-KKR method the 
dc-conductivity tensor {\boldmath$\sigma$} has been evaluated on the basis of 
the  Kubo-St\v{r}eda equation:\cite{Str82}
%
\begin{eqnarray}
\label{eq:streda}
\sigma_{\rm xy}& =& \frac{\hbar }{4\pi N\Omega}
                 {\rm Trace}\,\big\langle \hat{j}_{\rm x} (G^+-G^-) \hat{j}_{\rm y}  G^-  
                      \nonumber   \\  
&&  \qquad \qquad \quad 
-  \hat{j}_{\rm x} G^+\hat{j}_{\rm y}(G^+-G^-)\big\rangle_{\rm c} \nonumber \\
&& + \frac{e}{4\pi i N\Omega} {\rm Trace}\,
\big\langle (G^+-G^-)(\hat{r}_{\rm x}\hat{j}_{\rm y} - \hat{r}_{\rm y}\hat{j}_{\rm x}) \big\rangle_{\rm c} 
\label{eq:bru} 
\; .
\end{eqnarray}
%
%
See for example Ref.~\onlinecite{KCE15} concerning the implementation 
of this expression.

The optical conductivity tensor 
{\boldmath$\sigma$}$(\omega)$  for finite frequencies, on the other hand,
 has been determined using the SPR-LMTO method and the
 standard expression for the absorptive part of the diagonal
 and off-diagonal tensor elements,
 $\sigma_{\lambda\lambda}^1(\omega)$ and 
 $\sigma_{\lambda\lambda'}^2(\omega)$, respectively:\cite{WC74,Ebe96,AYPN99} 
%
%
\begin{eqnarray} 
\sigma_{\lambda\lambda}^1(\omega) & = & \frac{\pi e^2}{\hbar\omega m^2 V}
\sum_{
\begin{array}{c} 
j'\vec{k}\,{\rm occ.}\\
j\vec{k}\,{\rm unocc.}\end{array}
}
 | \Pi_{jj'}^\lambda |^2 \delta(\omega-\omega_{jj'}) \label{SIGDIA1} \\
%
%
\sigma_{\lambda\lambda'}^2(\omega) & = & \frac{\pi e^2}{\hbar\omega m^2 V}
\sum_{
\begin{array}{c} 
j'\vec{k}\,{\rm occ.}\\
j\vec{k}\,{\rm unocc.}\end{array}
}
\Im\left(\Pi_{j'j}^\lambda\Pi_{jj'}^{\lambda'}
\right)\delta(\omega-\omega_{jj'})\nonumber\\ \label{SIGOFF2}
\end{eqnarray}
%
%
with the matrix elements
 $\Pi_{j'j}^\lambda = \langle \phi_{j'\vec{k}} | {\cal H}_\lambda |\phi_{j\vec{k}}\rangle $
 of the Bloch states $ |\phi_{j\vec{k}}\rangle$
  and their energy difference
$\omega_{jj'}=E_{j\vec k} - E_{j'\vec k}$.
The dispersive part was determined in a second step by means of a Kramers-Kronig transformation.
With the full tensor {\boldmath$\sigma$}$(\omega)$ 
 available, the Kerr rotation angle  $ \theta_{\rm K} $ 
 was obtained from the standard expression:\cite{Ebe96,AYPN99}
%
%
\begin{eqnarray}
                                                           \label{phiKsig}
\theta_{\rm K} & \simeq & \Re 
\left( \frac{\sigma_{\rm xy}(\omega)}{\sigma_{\rm xx}(\omega)\sqrt{1-\frac{4\pi i}
{\omega}\sigma_{\rm xx}(\omega)}} 
\right)
\;,
\end{eqnarray}
%
%
that clearly shows that the Kerr rotation -- as the Kerr ellipticity --
is directly connected with the off-diagonal
element $\sigma_{\rm xy}(\omega)$ of the conductivity tensor.

Finally, the X-ray absorption coefficients $\mu_{\lambda}(\omega)$
 for polarization $\lambda$ have been calculated using the 
 SPR-KKR Green function method on the basis of the expression:\cite{Ebe96,EKM11}
%
%
\begin{equation}
\label{MXGFUN} 
\mu_{\lambda} (\omega )\propto
\sum_{\scriptstyle i\,{\rm occ}}
\langle \Phi_i|
{\cal H}_{\lambda}^{\dagger}\,\Im G^{+}(E_i+\hbar \omega )\,{\cal H}_{\lambda}
|\Phi_i\rangle \,\theta (E_i+\hbar \omega -E_{\rm F}) \;,
\end{equation}
%
%
where the functions $\Phi_i$ represent the probed
core states $i$ at energy $E_i$ and $E_{\rm F}$ is the Fermi energy. 

The corresponding XMCD signal 
$\Delta \mu(\omega)=\frac { 1 } { 2 } (\mu_{+}(\omega)- \mu_{-}(\omega) )$
is defined as the difference in absorption for left and right circularly polarized radiation.
 Expressing the absorption coefficient 
 $\mu_{\lambda}(\omega)$ in terms of the absorptive part of the 
corresponding conductivity tensor element:\cite{Ebe96}
%
%
\begin{eqnarray}
\label{eq:musig}
\mu_{\lambda}(\omega) & = & 
 \frac{4\pi}{c}\sigma_{\lambda}^1(\omega)\;,
\end{eqnarray}
%
%
with $\sigma_{\lambda}$ given by
%
%
\begin{eqnarray}
\label{eq:sigSCtrafo}
\sigma_{\pm}(\omega)= \sqrt{1/2 } \, (\sigma_{\rm xx}(\omega) \pm i \sigma_{\rm xy}(\omega)  ) \;,
\end{eqnarray}
%
%
one sees immediately that the occurrence of the off-diagonal element 
$\sigma_{\rm xy}(\omega) $ implies the occurrence of a XMCD signal $\Delta \mu(\omega)$.

\section{Cubic M\lowercase{n}$_3$I\lowercase{r} \label{sec:Mn3Ir}}

\begin{figure}
 \begin{center}
 \includegraphics[angle=0,width=0.45\linewidth,clip]{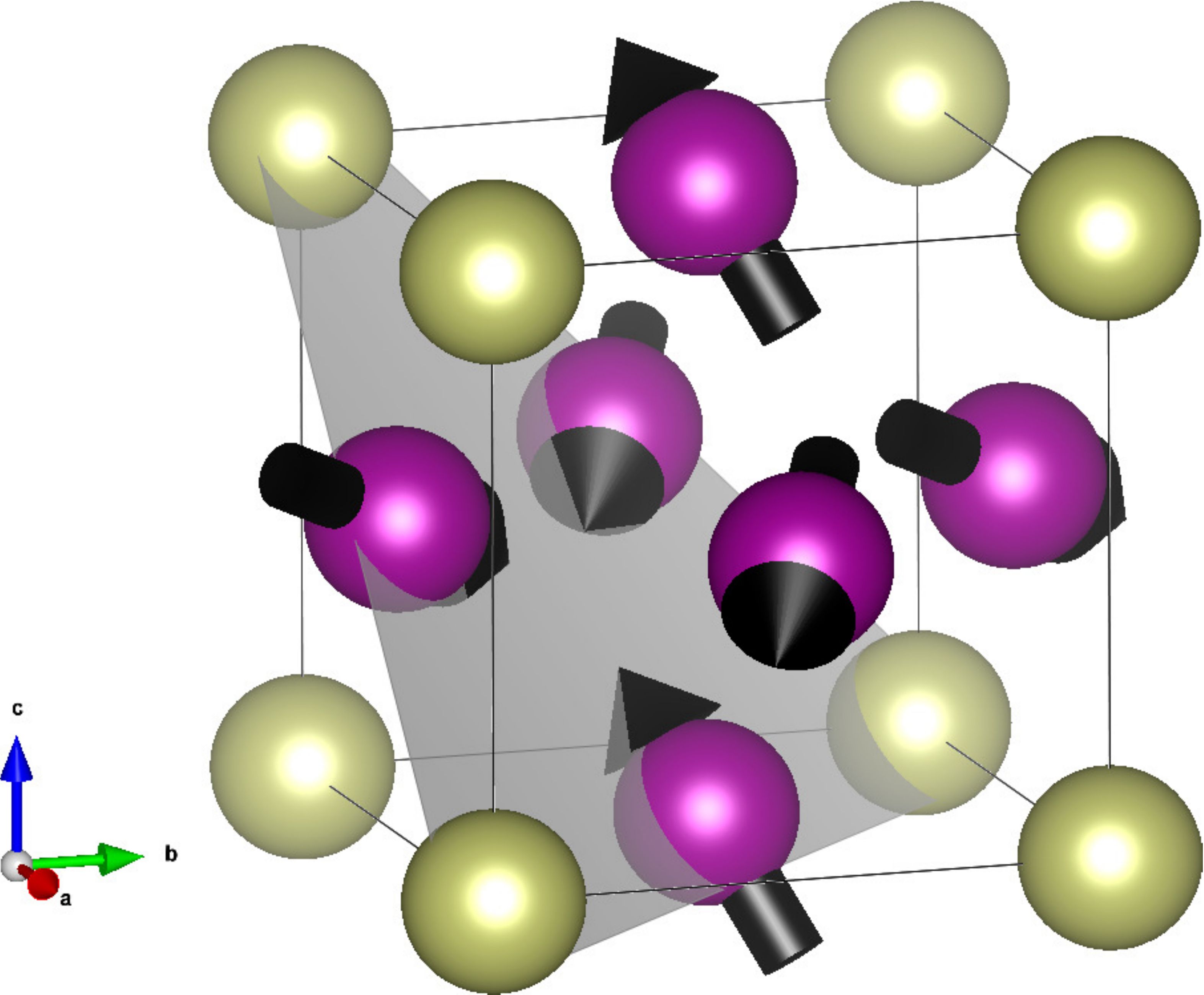}%
 \includegraphics[angle=0,width=0.55\linewidth,clip]{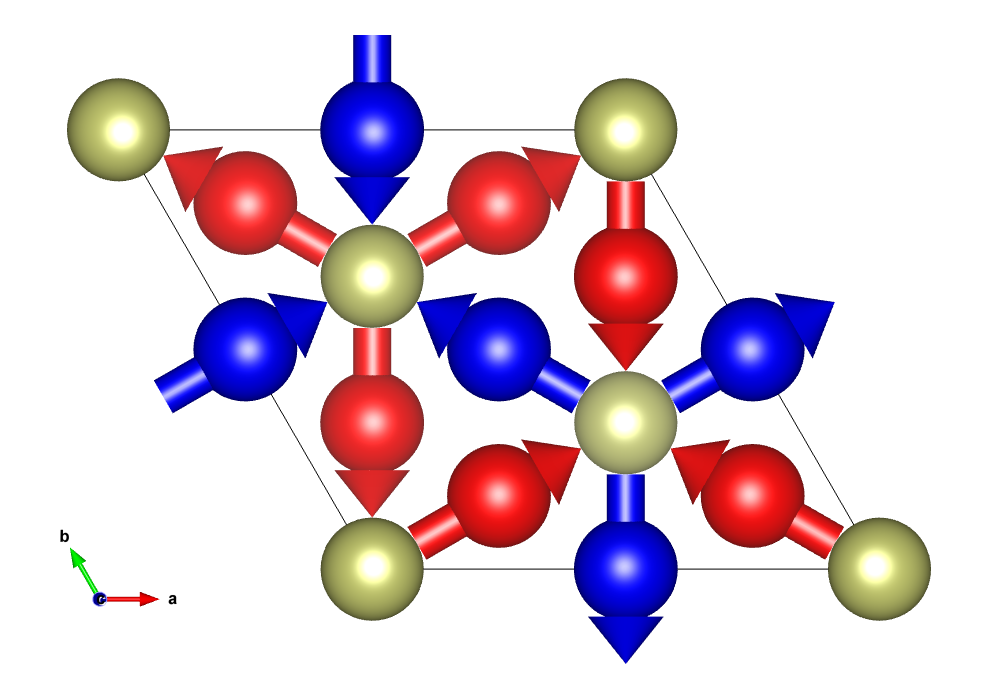}
   \caption{\label{FIG:Mn3Ir} Left: Cubic unit cell of Mn$_3$Ir in the 
	 triangular antiferromagnetic structure. Ir atoms are colored in gold
	 (light gray), Mn atoms in purple (dark gray). Black arrows indicate
	 the direction of the magnetic moments on the Mn sites, the gray-shaded
	 surface marks the (111) plane. Right: View along [111] on the Kagome
	 planes, Mn atoms and moments in alternating layers are colored in red
	 and blue.\cite{VESTA}}
\end{center}
\end{figure}

The structure and magnetic configuration of Mn$_3$Ir is given in Fig.\
\ref{FIG:Mn3Ir} showing a non-collinear antiferromagnetic spin arrangement.
%
%
The Mn moments are lying in the $\{111\}$ planes of the underlying Cu$_3$Au
(L1$_2$) lattice and are oriented along the $\langle112\rangle$ directions,
forming a Kagome lattice of corner-sharing triangles. Calculated values for spin 
and orbital moments are given in Table~\ref{tb:mmoms}.
%
\begin{table}[h]
\begin{tabular}{cccc}
 & $\mu_{\rm spin} (\mu_{\rm B})$ & $\mu_{\rm orb.} (\mu_{\rm B})$ & $\hat{\mu}$\\
\toprule
Mn & 2.849 & 0.066 & $\langle112\rangle$ \\
Ir & -0.001 & -0.033 & $[111]$ \\
tot. & -0.001 & -0.033 & $[111]$ \\
\end{tabular}
\caption{\label{tb:mmoms} Spin and orbital magnetic moments of Mn$_3$Ir, atomic 
type-resolved as well as total values are given in units of $\mu_{\rm B}$. Their 
directions are given in the last column.}
\end{table}
%
Obviously, the magnetic moment of the Mn atoms is dominated by the spin 
contribution. For Ir on the other hand the very small induced moment is 
primarily due to its orbital part.

The magnetic structure leads to the magnetic space group $R\overline{3}m'$
that has to be considered instead of the space group $Pm\overline{3}m$ (221)
of the Cu$_3$Au structure. The magnetic point group corresponding to
$R\overline{3}m'$ is $\overline{3}m'$ and independent of the definition used
for the  magnetic Laue group, this implies the following shape of the
conductivity tensor:
%
\begin{equation}
  \label{eq:sigma-Mn3Ir}
	\underline{\mbox{\boldmath$\sigma$}}(\omega)
  =
  \left(
    \begin{array}{ccc}
      \sigma_{xx}(\omega) & \sigma_{xy}(\omega) & 0 \\
     -\sigma_{xy}(\omega) & \sigma_{xx}(\omega) & 0 \\
      0                   & 0                   & \sigma_{zz}(\omega)
    \end{array}
  \right)
  \; .
\end{equation}
%
Here the indices refer to a coordinate system that is conform with the symmetry
of the system having the z-axis  along the conventional [111]-direction  of the
cubic Cu$_3$Au (L1$_2$) lattice, while the x- and y-axes lie  in the
\{111\}-plane. 
 Strictly spoken, our self-consistent calculations give the Mn magnetic
moment slightly tilted from the \{111\} plane by
$\Delta\theta = 0.03^\circ$. This small tilting does not change the
symmetry of the system and has only little influence on the transport
properties. For that reason it was not taken into account when performing
the transport calculations. 

Obviously, {\boldmath$\sigma$}$(\omega)$ has exactly the same shape as any fcc
or bcc ferromagnetic material with the magnetization along the z-axis that
coincides with the [001]-direction.\cite{Ebe96} This implies that any
gyro-magnetic and magneto-optical phenomena occurring for this well-known
situation will also be present for the non-collinear antiferromagnet Mn$_3$Ir.
For the corresponding spin conductivity tensor shapes see Ref.~\onlinecite{SKWE15a}.


\iftransport
\subsection{Anomalous and spin Hall effect}

In the DC-limit ($\omega=0 $) the conductivity tensor 
{\boldmath$\sigma$}$(\omega)$  given by Eq.~\eqref{eq:sigma-Mn3Ir} becomes real. 
For pure systems the diagonal elements representing longitudinal conductivity 
diverge for $T = 0$~K, while the off-diagonal anomalous Hall conductivity (AHC) 
$\sigma_{\rm xy} = - \sigma_{\rm yx}$ stays finite. Chen \emph{et al.} 
\cite{CNM14} calculated the AHC using an expression in terms of the  Berry 
curvature. Their result  $\sigma_{\rm AH} =218$\, ($\Omega\cdot$cm)$^{-1}$  is 
comparable in size with that for the ferromagnets Fe, Co, and Ni. In the present 
work the Kubo-St\v{r}eda linear response formalism\cite{Str82,LKE10b,KCE15} was 
used, that for pure systems is completely equivalent to the Berry curvature 
approach.

The values obtained for the intrinsic anomalous and spin Hall conductivities at 
$T = 0$\,K in pure Mn$_3X$ with $X = $ Ir, Pt, and Rh using the Kubo-St\v{r}eda 
formula are collected in Table~\ref{tb:XHC_cMn3X}.
%
\begin{table}[h]
\begin{tabular}{c|c|c}
         & AHC & SHC \\
\toprule
Mn$_3$Rh & -85       & -125 \\
Mn$_3$Ir & -280 & -230 \\
Mn$_3$Pt & -360      & -250 \\
\end{tabular}
\caption{\label{tb:XHC_cMn3X} Anomalous and spin Hall conductivities, AHC = 
$\sigma_{xy}$ and SHC = $\sigma_{xy}^z$, respectively, for pure cubic Mn$_3X$ 
compounds with $X = $ Rh, Ir, and Pt in units of ($\Omega\cdot$cm)$^{-1}$. The $z$ direction 
corresponds to the cubic [111] direction perpendicular to the Kagome planes. Results 
are obtained using the Kubo-St\v{r}eda formula.}
\end{table}
%
For the AHC in Mn$_3$Ir reasonable agreement with previous results by 
Chen~\emph{et al.}\cite{CNM14} ($\vert218\vert$ ($\Omega\cdot$cm)$^{-1}$) and Zhang~\emph{et 
al.}\cite{ZSY+17} (-312 ($\Omega\cdot$cm)$^{-1}$) is achieved. The clear trend of increasing AHC as 
well as SHC with atomic number is however in disagreement with the findings of 
the latter authors.

%
%
\begin{figure}[h]
 \begin{center}
 \includegraphics[angle=0,width=0.8\linewidth,clip]{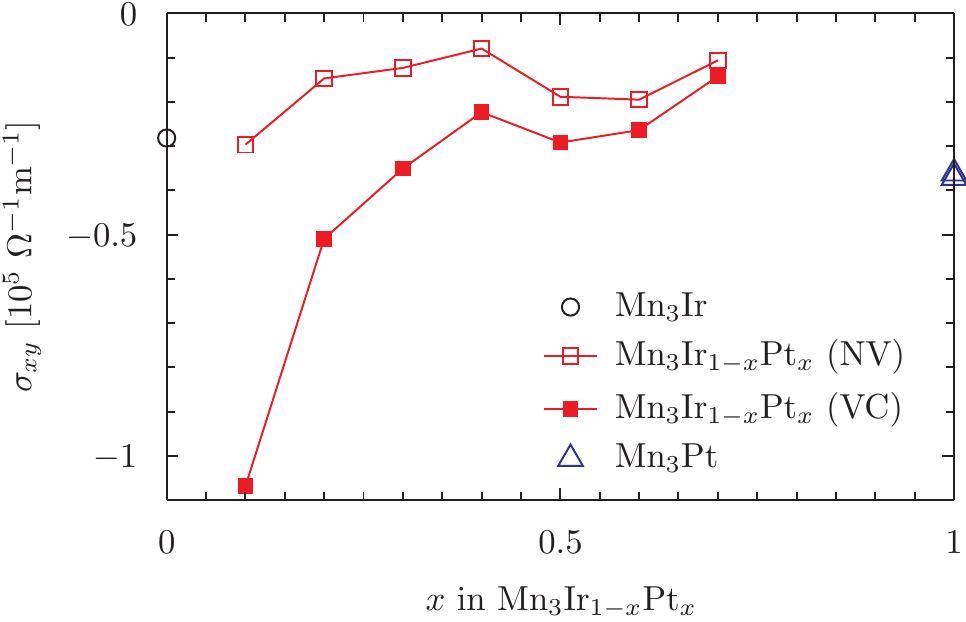}
 \includegraphics[angle=0,width=0.8\linewidth,clip]{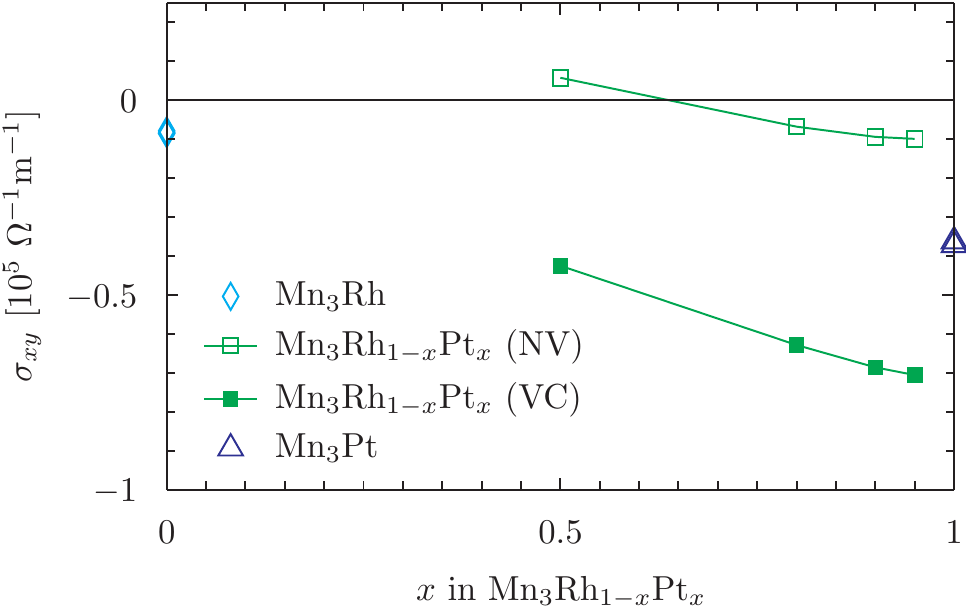}
 \includegraphics[angle=0,width=0.8\linewidth,clip]{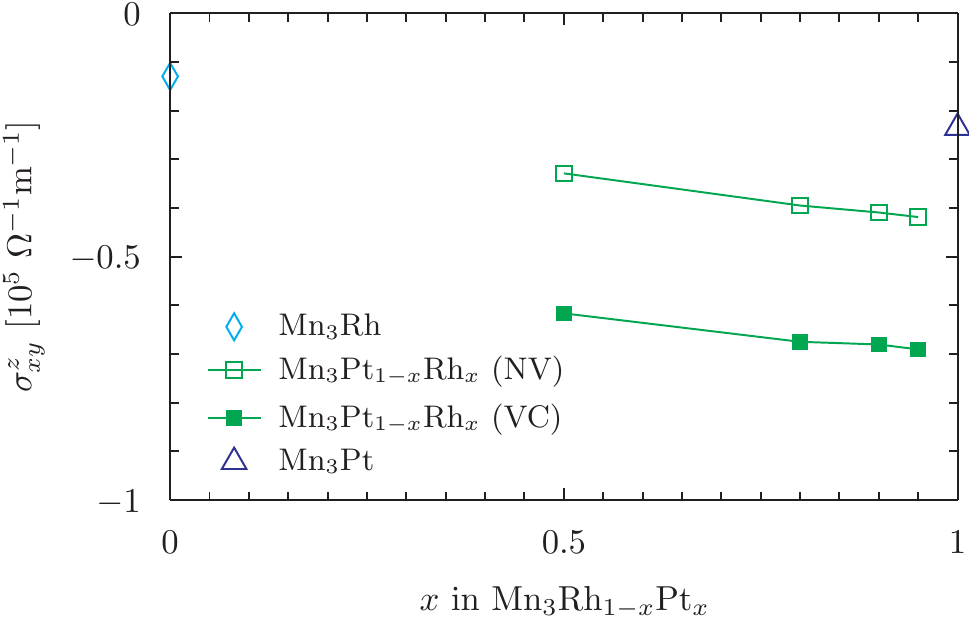}
  \caption{\label{fig:AHE-Mn3Pt-alloys} Anomalous Hall conductivity of 
Mn$_3$Ir$_{1-x}$Pt$_x$ (top) and Mn$_3$Rh$_{1-x}$Pt$_x$ (middle) as well as spin 
Hall conductivity of Mn$_3$Rh$_{1-x}$Pt$_x$ alloys (bottom) as a function of 
concentration $x$. Open symbols are results excluding vertex corrections (NV) 
and filled symbols such including them (VC), see text for details. The intrinsic 
values of the pure compounds are shown in addition. All results were obtained 
using the Kubo-St\v{r}eda formula.}
\end{center}
\end{figure}
%
Figure~\ref{fig:AHE-Mn3Pt-alloys} shows results for the concentration dependence 
of the AHC in Mn$_3$Ir$_{1-x}$Pt$_x$ (top) as well as for AHC and SHC in 
Mn$_3$Rh$_{1-x}$Pt$_x$ (middle and bottom, respectively). While for 
$\sigma_{xy}$ in Mn$_3$Ir$_{1-x}$Pt$_x$ the usual divergence of the values 
including vertex corrections is found, followed by a non-trivial concentration 
dependence, in Mn$_3$Rh$_{1-x}$Pt$_x$ the impact of these is large for the 
whole investigated range. The spin Hall conductivity in this system behaves very 
similar to its charge counterpart. The fact that both conductivities without 
vertex corrections (NV) do not clearly converge towards the intrinsic values 
when approaching the dilute limits $x \rightarrow 0/1$ still has to be clarified. 
For AHC and SHC in Mn$_3$Rh$_{1-x}$Pt$_x$ an almost linear increase with 
increasing concentration of the heavier alloy partner Pt can be observed, the 
intrinsic contribution (NV) in particular for the AHC however appears to behave 
again non-trivially for intermediate concentrations.

\fi
\color{black}

\subsection{Magneto-optical Kerr effect (MOKE)}

\begin{figure}
 \begin{center}
 \includegraphics[angle=0,width=0.8\linewidth,clip]{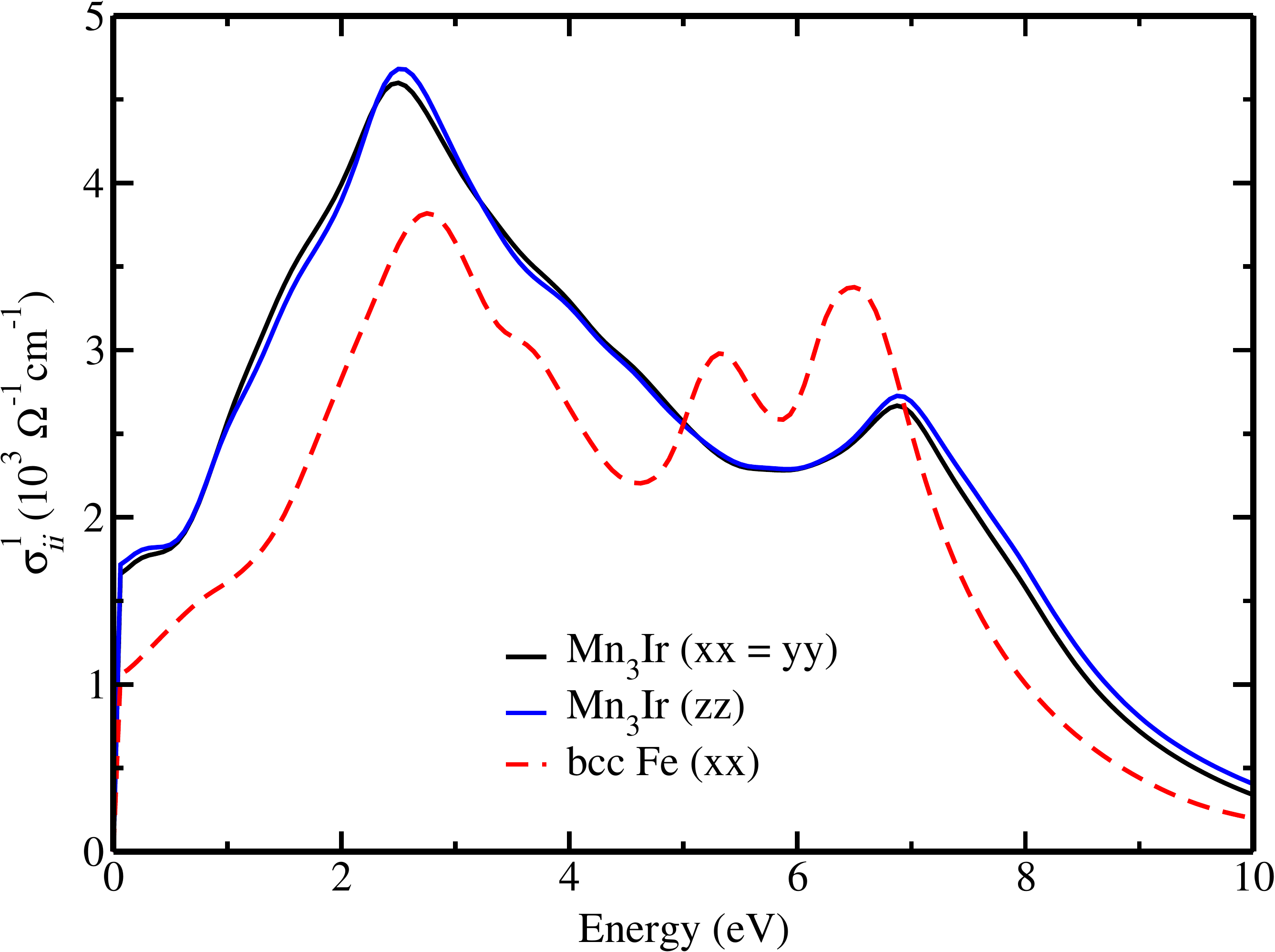}
 \includegraphics[angle=0,width=0.8\linewidth,clip]{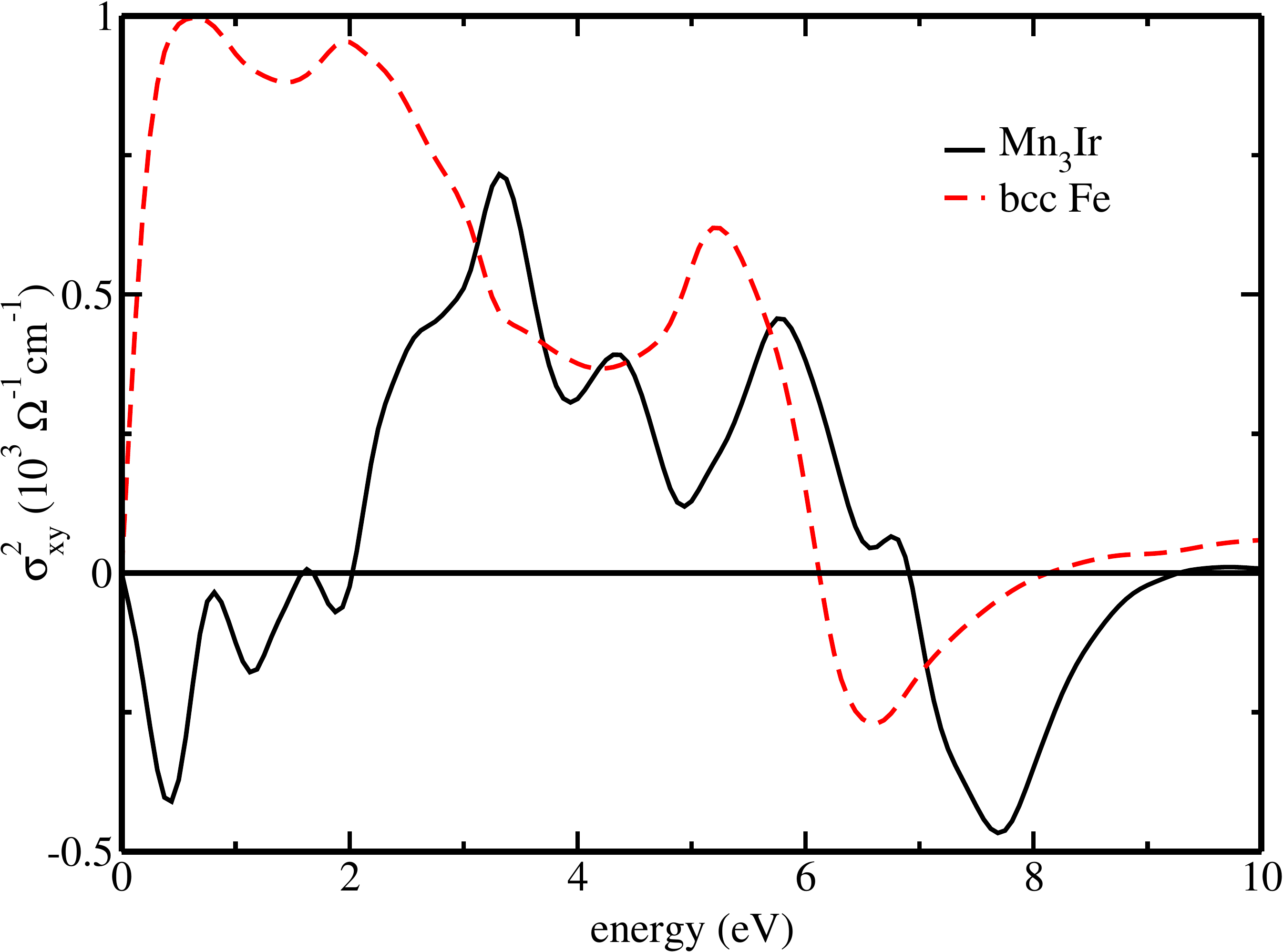}
 \includegraphics[angle=0,width=0.8\linewidth,clip]{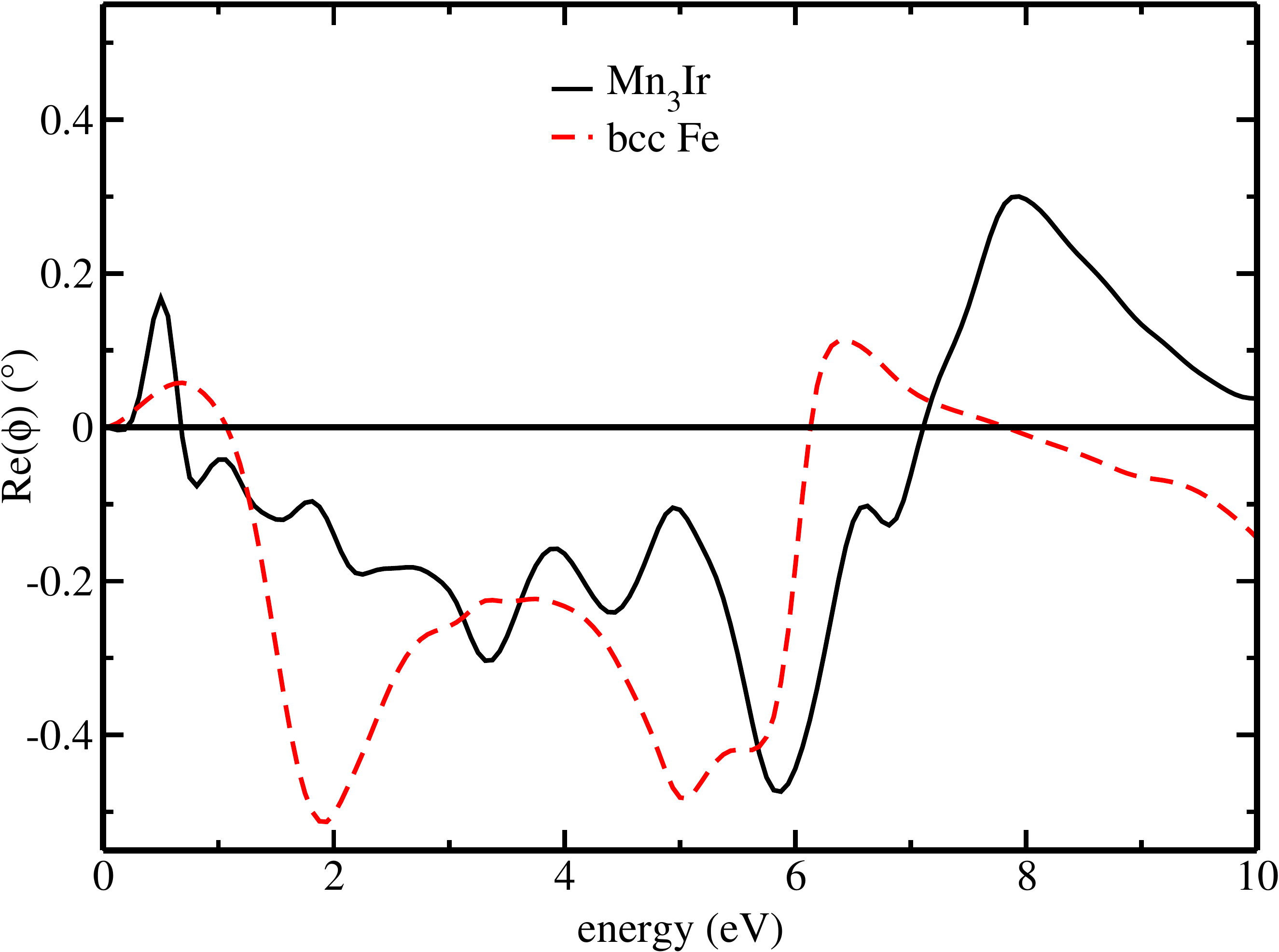}
  \caption{\label{fig:MO-Mn3Ir-vs-bccFe} Absorptive part of the optical 
conductivity tensor elements $\sigma_{\rm xx}^{1}$, $\sigma_{zz}^{1}$ (top) and 
$\sigma_{\rm xy}^{2}$ (middle) of Mn$_3$Ir (full lines) together with the Kerr 
rotation angle $\theta_{\rm K}$ (bottom). In addition, corresponding results for 
ferromagnetic bcc-Fe are given (dashed lines).}
\end{center}
\end{figure}

For finite frequencies the absorptive part of the corresponding optical 
conductivity tensor {\boldmath$\sigma$}$(\omega)$ has been calculated in the 
energy regime $\hbar \omega = 0\,- \, 10$~eV using the fully relativistic LMTO 
band structure method.\cite{Ebe88,PY-LMTO,AHY04} 
Figure~\ref{fig:MO-Mn3Ir-vs-bccFe} (top) shows the corresponding real parts of 
the optical conductivity, $\sigma^1_{ii}(\omega)$.
%
%
A reasonable qualitative agreement with earlier theoretical work by Feng~\emph{et 
al.}\cite{FGZ+15a} is found. The small difference between 
$\sigma^1_{xx}(\omega)=\sigma^1_{yy}(\omega)$ and $\sigma^1_{zz}(\omega)$ 
obviously reflects the anisotropy of the system due to the underlying lattice 
structure as well as the magnetic ordering. Accordingly it will give rise to 
corresponding magneto-optical phenomena.\cite{Ebe96} As a reference, 
Fig.~\ref{fig:MO-Mn3Ir-vs-bccFe}  (top) gives also corresponding results for 
ferromagnetic bcc-Fe. In this case the anisotropy, i.e., the difference between
$\sigma^1_{xx}(\omega)$ and $\sigma^1_{zz}(\omega)$, is only due to the magnetic 
ordering. As it is much less pronounced than for Mn$_3$Ir only 
$\sigma^1_{xx}(\omega)$ is given.

Figure~\ref{fig:MO-Mn3Ir-vs-bccFe}  (middle) shows the imaginary part of  the
off-diagonal transverse optical conductivity, $\sigma^2_{\rm xy}$, that is
the counterpart to the AHC and that in particular gives rise to the polar
Kerr rotation. Interestingly, $\sigma^2_{\rm xy}$ for Mn$_3$Ir  is in the
same order of magnitude as for ferromagnetic Fe. Accordingly one finds
$\theta_{\rm K}$ of Mn$_3$Ir and Fe to be comparable in magnitude (see bottom
panel of Fig.~\ref{fig:MO-Mn3Ir-vs-bccFe}). Again reasonable qualitative agreement
with Ref.~\onlinecite{FGZ+15a} is found.\\


\subsection{X-ray absorption spectroscopy \label{ssec:XAS-Mn3Ir}}

\begin{figure}
 \begin{center}
 \includegraphics[angle=0,width=0.8\linewidth,clip]{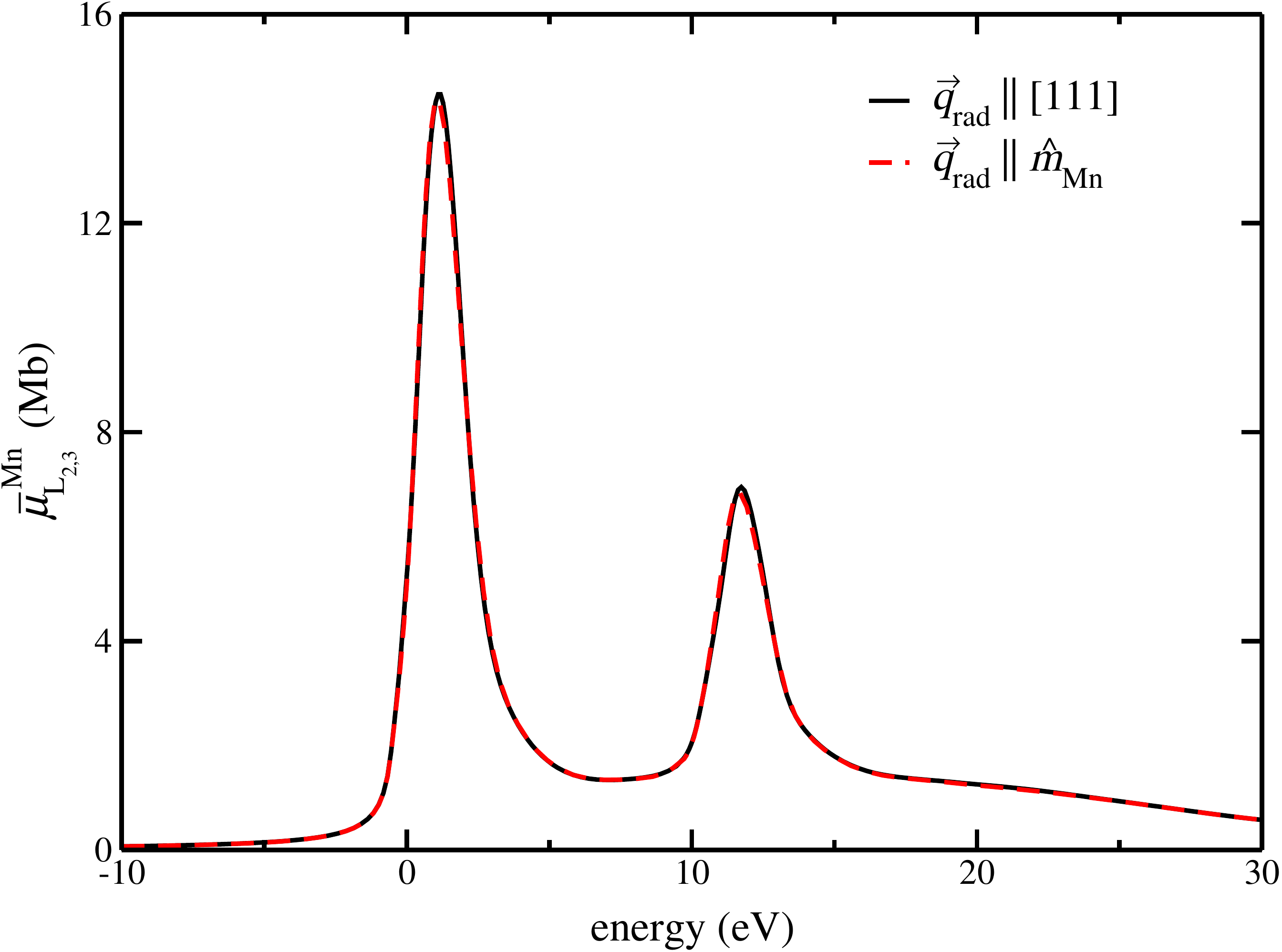}
 \includegraphics[angle=0,width=0.8\linewidth,clip]{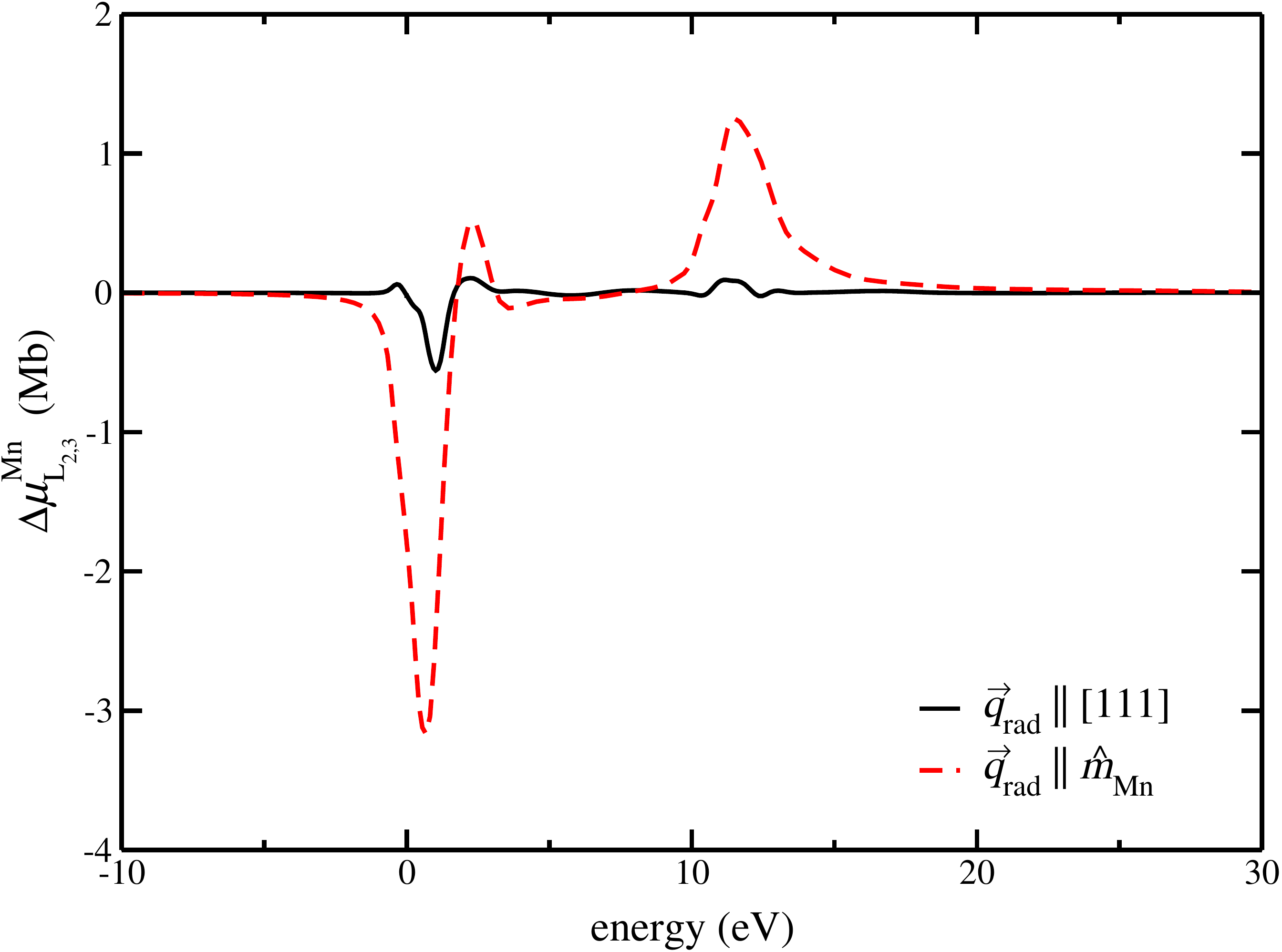}
  \caption{\label{fig:XAS-Mn3Ir-111VSpolar} Total absorption coefficient $\bar\mu$
  at the L$_{2,3}$-edge of a single Mn atom of non-collinear antiferromagnetic Mn$_3$Ir 
  (top) and corresponding circular dichroism spectra $\Delta\mu$ (bottom), both 
  for incidence $\vec{q}_{\rm rad}$ perpendicular (along [111]) and polar w.r.t.\ 
  the magnetization direction.}
\end{center}
\end{figure}

Fig.~\ref{fig:XAS-Mn3Ir-111VSpolar} (upper panel) shows results of calculations
for the X-ray absorption coefficient $ \bar \mu = \frac{ 1 } { 2 }
(\mu_{+}+\mu_{-} )$ at the Mn L$_{2,3}$-edges for polarization-averaged
radiation. These spectra show the typical L$_{2, 3}$-edges of a 3d transition
metal shifted against each other by the spin-orbit splitting of the 2p core
states. The two curves shown have been obtained for the radiation wave vector
$\vec q_{ \rm rad } $ along the [111]-direction and along the direction $ \hat
m_{\rm Mn}$ of the magnetic moment of one of the three equivalent Mn-atoms in
the unit cell. Similar to $\sigma_{\rm xx}^{1}$ and  $\sigma_{zz}^{1}$ in the
optical regime discussed above the difference  is quite small, i.e., only a weak
anisotropy occurs. The lower panel of Fig.~\ref{fig:XAS-Mn3Ir-111VSpolar} gives
the corresponding XMCD curves $\Delta \mu$.
%
%
For the polar geometry $\vec q_{ \rm rad } \parallel \hat m_{\rm Mn}$ with the 
artificial restriction to one of the Mn sites the highest XMCD signal can be 
expected. Indeed for the L$_3$-edge about 20~\% is found for the ratio $\Delta 
\mu/\bar \mu$. This is the typical order of magnitude found in ferromagnetic 
transition metals.\cite{Ebe96} Considering instead the geometry $\vec q_{ \rm 
rad } \parallel [111]$ with the X-ray beam perpendicular to the Mn magnetic 
moments, no XMCD signal is expected following the standard 
arguments.\cite{Ebe96} In contrast to this, Fig.~\ref{fig:XAS-Mn3Ir-111VSpolar} 
(lower panel) clearly shows that there is indeed a finite XMCD present as one 
had to expect on the basis of Eqs.~\eqref{eq:musig}, \eqref{eq:sigSCtrafo} and 
\eqref{eq:sigma-Mn3Ir}. In addition one has to emphasize that the individual 
XMCD for the three Mn sites are identical, i.e., they do not compensate each 
other as in the case of $\vec q_{ \rm rad } \parallel \hat m_{\rm Mn}$ but add 
up. Comparing the two XMCD spectra in Fig.~\ref{fig:XAS-Mn3Ir-111VSpolar} one 
notes that  $\Delta \mu$ for $\vec q_{ \rm rad } \parallel [111]$ is very 
similar in shape to that for $\vec q_{ \rm rad } \parallel \hat m_{\rm Mn}$ but 
about one order of magnitude smaller in amplitude. Nevertheless, this implies 
that it should be possible to detect this XMCD signal in experiment provided one 
domain dominates in the regime exposed to the X-ray beam (see also comment by 
Chen \emph{et al.}\cite{CNM14}).\\
%
\begin{figure}
 \begin{center}
 \includegraphics[angle=0,width=0.8\linewidth,clip]{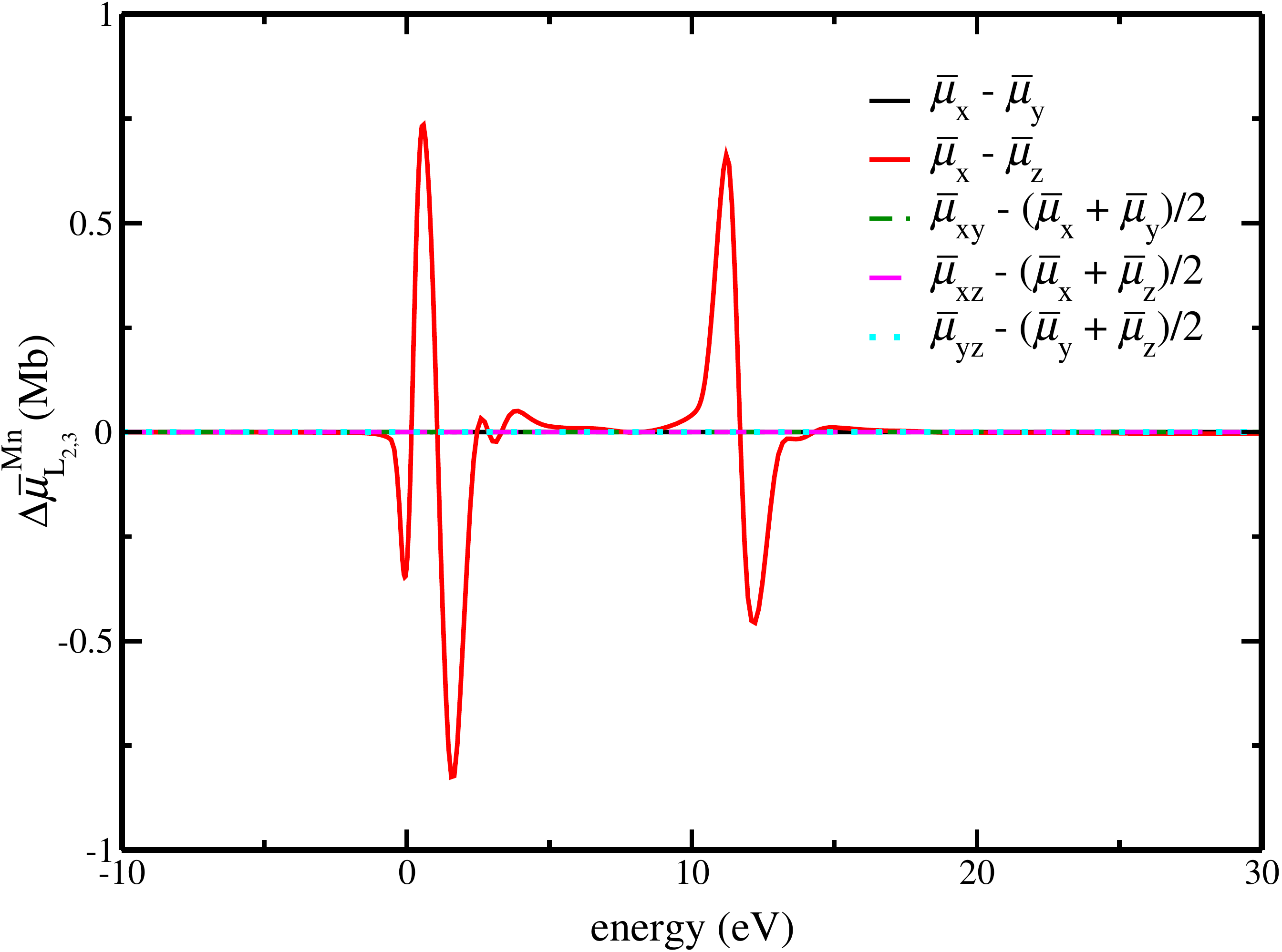}
 \includegraphics[angle=0,width=0.8\linewidth,clip]{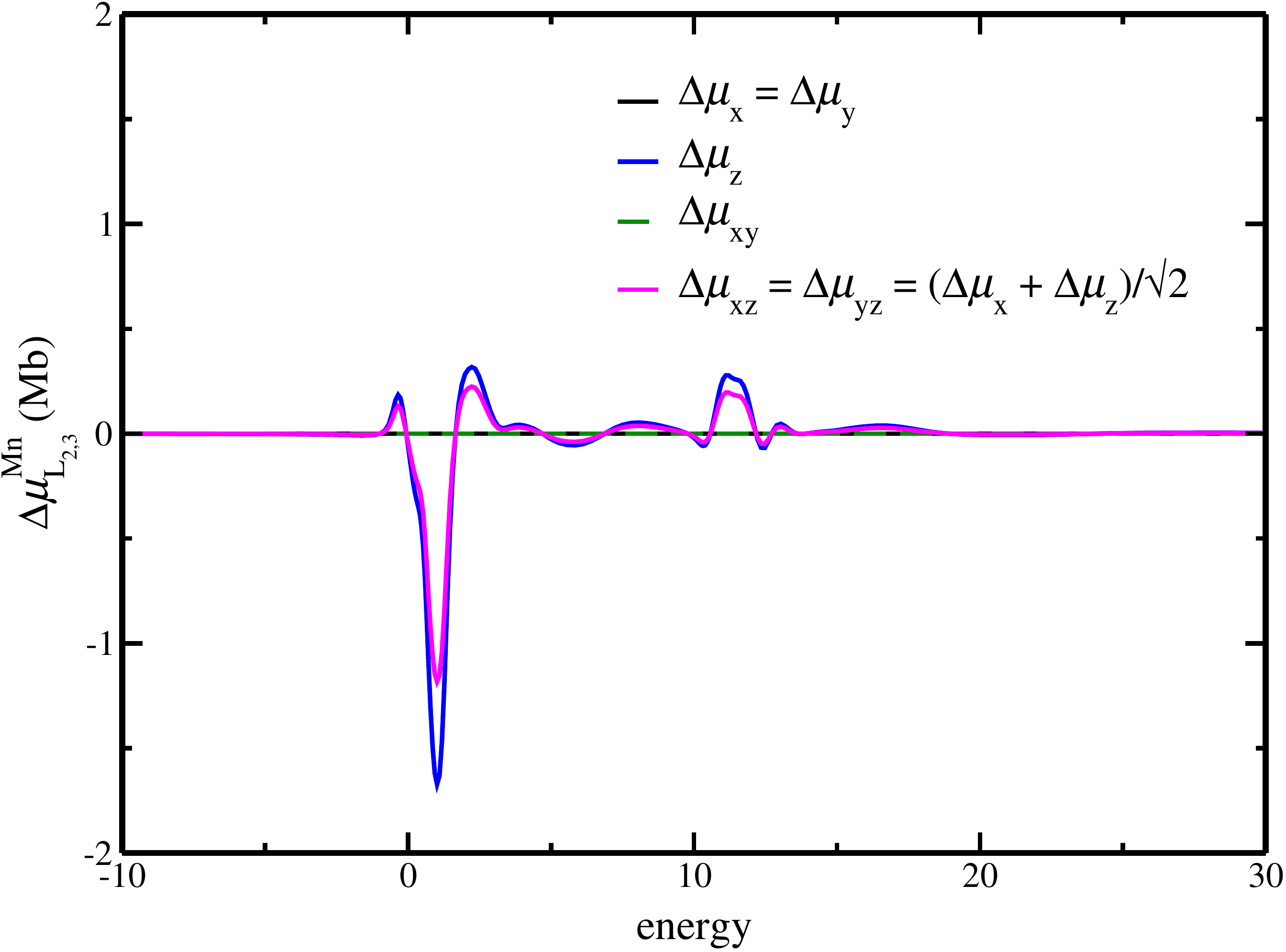}
  \caption{\label{fig:XAS-Mn3Ir} Top: Differential polarization-averaged X-ray 
absorption spectra  $\Delta\bar\mu$ comparing incidence along high-symmetry 
directions in Mn$_3$Ir. Bottom: X-ray magnetic circular dichroism (XMCD) spectra 
$\Delta\mu$ in Mn$_3$Ir for incidence along the same directions as in the top 
panel.}
 \end{center}
\end{figure}
%

In the top panel of Figure~\ref{fig:XAS-Mn3Ir} the difference in total
absorption $\Delta\bar\mu = \bar\mu_i - \bar\mu_j$ with $\bar\mu_i =
\frac{1}{2}(\mu_{i,+} + \mu_{i,-})$ for incidence along pairs of high symmetry
directions ($i,j = \rm x,y,z$) and along intermediate directions ($i = \rm
xy,xz,yz$) compared to the corresponding linear combinations of x, y, z is shown.
Note that in contrast to the conventional X-ray magnetic linear dichroism, here the 
polarization of the incoming light is not rotated.
The Cartesian directions x and y were chosen to be in the (111) plane (indicated
in gray in Figure~\ref{FIG:Mn3Ir}), i.e., corresponding to the crystallographic
directions [11$\bar 2$] (x) and [$\bar 1$10] (y) of the cubic unit cell, while z is parallel to the
[111] direction (the space diagonal of the cubic unit cell shown in
Fig.~\ref{FIG:Mn3Ir}). As can be seen, the absorption is isotropic in the (111)
plane ($\bar\mu_{\rm x} = \bar\mu_{\rm y}$), but different for incidence perpendicular to it,
i.e., along the [111] direction ($\bar\mu_{\rm z} \neq \bar\mu_{\rm x}$). This agrees
with the diagonal elements of the tensor in Eq.~\eqref{eq:sigma-Mn3Ir}, since the first
two elements are identical, $\sigma_{\rm xx} = \sigma_{\rm yy}$, and
different from the third ($\sigma_{\rm zz}$). Furthermore the absorption for
intermediate directions (xy, xz and yz) can be described by a linear combination
of the absorption coefficients along Cartesian axes, such as e.g. $\bar\mu_{\rm
xz} = \frac{1}{2}(\bar\mu_{\rm x} + \bar\mu_{\rm z})$, i.e., no symmetric
off-diagonal elements are present in the absorption tensor. In the lower panel
of Fig.~\ref{fig:XAS-Mn3Ir} the circular dichroism for several directions of
incidence is compared. Again the results are in agreement with the tensor shape
of Eq.~\eqref{eq:sigma-Mn3Ir}, inasmuch that the only linearly independent
non-zero anti-symmetric tensor element is $\Delta\mu_{\rm z}$, corresponding to
$\sigma_{\rm xy} = -\sigma_{\rm yx}$.

\section{Hexagonal M\lowercase{n}$_3$G\lowercase{e} \label{sec:Mn3Ge}}

\subsection{Magnetic structure and symmetry}

\begin{figure}[ht]
 \begin{center}
\vspace*{-0.375cm}
 \includegraphics[angle=0,width=0.9\linewidth,clip]{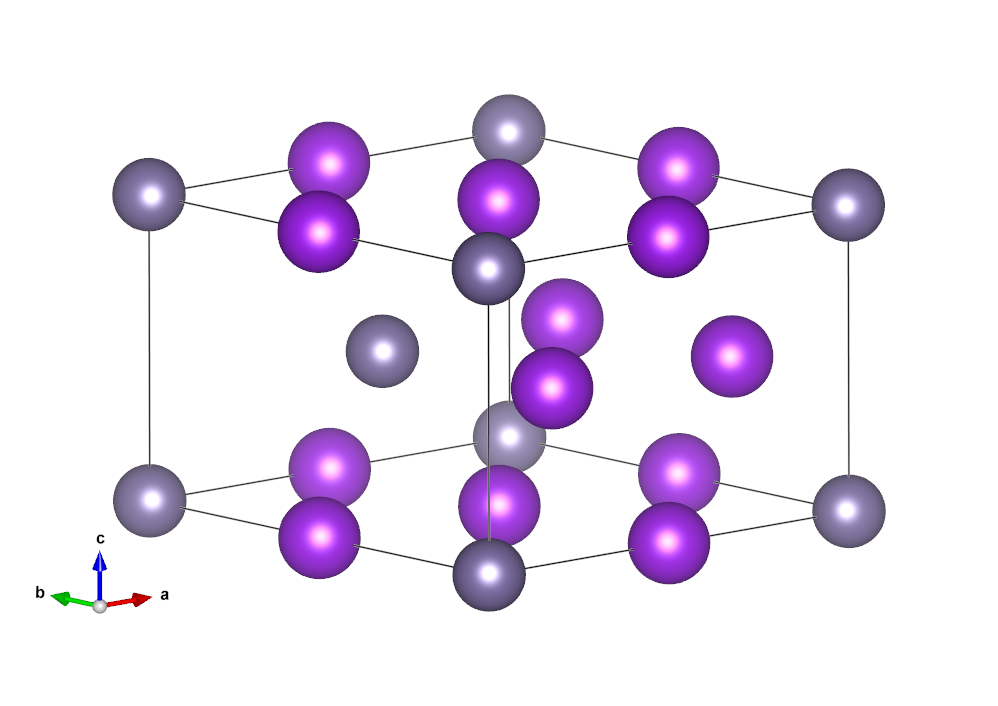} 
\vspace*{-0.5cm}
\end{center}
   \caption{\label{fig:Mn3Ge} Non-magnetic unit cell of Mn$_3$Ge 
in space group P6$_3$/mmc (194). Mn atoms on the Wyckoff positions $6h$ are 
colored in magenta (dark grey) and Ge atoms (Wyckoff positions $2h$) are colored 
in light gray.\cite{VESTA}}
\end{figure}

The hexagonal Mn$_3$Ge compound crystallizes, as its siblings Mn$_3$Sn and 
Mn$_3$Ga, in the D0$_{19}$ structure with space group $P6_3/mmc$ (194). The 
non-magnetic unit cell is shown in Fig.~\ref{fig:Mn3Ge} and will be labeled NM 
in the following. The Mn atoms on the Wyckoff positions $6h$ in the \{0001\} 
planes colored in magenta (dark gray) form triangular, so-called Kagome 
lattices, stacked alternatingly along the [0001] ($z$) direction. Ge 
atoms occupying the Wyckoff positions $2h$ are colored in light gray.
%
%

\begin{figure*}
\begin{center}
	\includegraphics[angle=0,width=0.9\linewidth,clip]{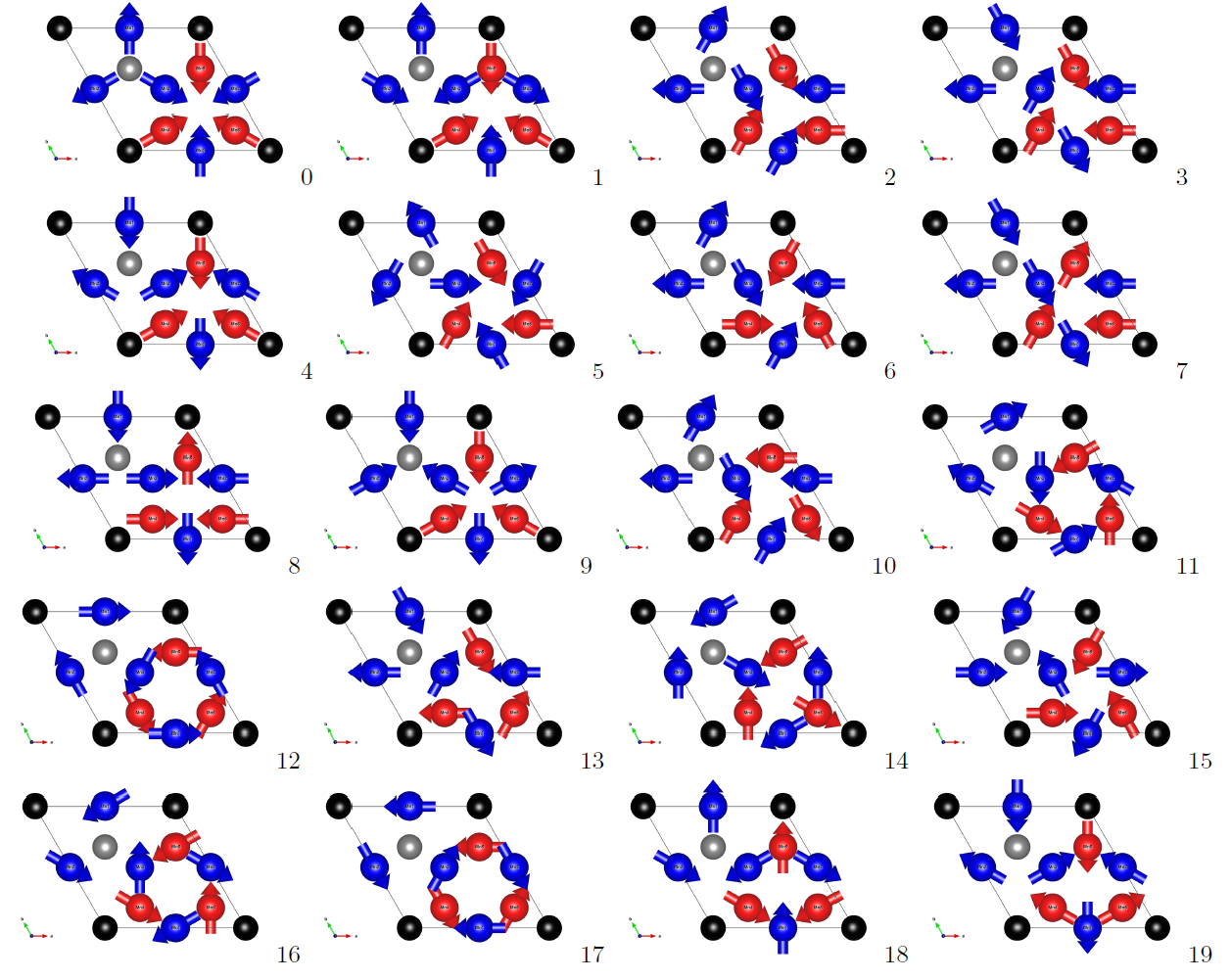}
\end{center}
   \caption{\label{fig:Mn3GePhAFMx} Antiferromagnetic structures 
of Mn$_3$Ge, AFM$i$ with $i = 0...19$, discussed in this contribution.\cite{VESTA}}
\end{figure*}

A number of non-collinear but coplanar antiferromagnetic alignments of the 
moments in the two Kagome sub-lattices have been discussed for Mn$_3$Ge and related 
compounds in the literature.\cite{KK70,NTY82,TY82,TYN83,ZYW+13,KF14a} These are 
collected in Figure~\ref{fig:Mn3GePhAFMx} together with further, hypothetical spin 
compensated structures.
%
%

The structure AFM0 has the same spin arrangement in both Kagome planes with the 
Mn moments pointing alternatingly towards the center of the corner-sharing 
triangles or away from it. The two sub-lattices are connected by a $6_3$ 
operation, i.e., a rotation around [0001] (z) by 60$^\circ$ combined with a 
translation along z by half of the corresponding lattice constant. This leads to 
a bipartite, globally chiral lattice with the non-magnetic Laue group $6/mmm1'$. 
It corresponds to Fig.~3(a) of Ref.~\onlinecite{ZYW+13}. The structures AFM1-4 
have been discussed by K\"ubler and Felser\cite{KF14a}, see Figs.~2, 3(a), 3(b), 
and 5 therein. The structure AFM3 is reported to be obtained by a 
self-consistent calculation starting from AFM2. An anomalous Hall effect has 
been predicted for AFM3 and AFM4 by these authors, as well as for an additional 
non-coplanar structure which will not be discussed herein. The structures AFM5 
and AFM6 are obtained from AFM0 by rotating all moments in the Kagome planes by 
30$^\circ$, counterclockwise and clockwise, respectively. AFM7, AFM8, and AFM10 
are hypothetical structures. AFM9 is obtained from AFM0 by reversing the moments 
in one Kagome layer (here the one in blue), leading to opposite chirality in the 
two planes and global inversion symmetry. The sub-lattices are connected by 
$6_3'$, i.e., a screw rotation in combination with time reversal. It corresponds 
to Fig.~3(b) of Ref.~\onlinecite{ZYW+13}. The structures AFM11 and AFM12 are 
60$^\circ$ and 90$^\circ$ clockwise rotations of all moments in AFM0. AFM13 and 
AFM14 have been proposed in Ref.~\onlinecite{KTN16}, the first experimental 
verification of an anomalous Hall effect in Mn$_3$Ge. AFM13 should furthermore 
correspond to the magnetic structure of Mn$_3$Sn in Ref.~\onlinecite{NKH15}, the 
very first reported observation of the AHE in a non-collinear antiferromagnet of 
the hexagonal Mn$_3X$ Heusler type. In AFM15, AFM16, and AFM17 
the moments are 30$^\circ$, 60$^\circ$, and 90$^\circ$ clockwise rotated w.r.t.\
AFM9. Finally, AFM18 and AFM19 are proposed in 
Refs.~\onlinecite{NFS+16,ZZFY17a,YSZ+17,ZZS+17,ZSY+17}, the latter of the two 
has been reported by Zhang \emph{et al.}\cite{ZYW+13} to be the most stable for 
Mn$_3X$ ($X = $ Ga, Sn, Ge) based on DFT calculations. One should stress here again, that all of the above magnetic structures 
are coplanar and fully spin-compensated, i.e., no weak ferromagnetism\cite{SK96} 
due to an out-of-plane rotation is taken into account here.

\begin{table}
\ifMPG
\begin{tabular}{c|c|c|c}
label & magn. space group & magn. point group & magn. Laue group\\
\toprule
NM    & $P6_3/mmc$    & $6/mmm1'$   & $6/mmm1'$  $(6221')$ \\
FM    & $P6_3/mm'c'$  & $6/mm'm'$   & $6/mm'm'$  $(62'2')$ \\
AFM0  & $P6_3/m'm'c'$ & $6/m'm'm'$  & $6/mmm1'$  $(6221')$ \\
AFM1  & $Am'm'2$      & $m'm'2$     & $m'm'm$    $(2'2'2)$ \\
AFM2  & $P\bar6'2c'$  & $\bar6'2m'$ & $6'/m'mm'$ $(6'22')$ \\
AFM3  & $Pm'$         & $m'$        & $2'/m'$    $ (2')$ \\
AFM4  & $Am'm'2$      & $m'm'2$     & $m'm'm$    $(2'2'2)$ \\
AFM5  & $P6_3/m'$     & $6/m'$      & $6/m1'$    $(61')$ \\
AFM6  & $P6_3/m'$     & $6/m'$      & $6/m1'$    $(61')$ \\
AFM7  & $Am'a'2$      & $m'm'2$     & $m'm'm$    $(2'2'2)$ \\
AFM8  & $Cm'c'm'$     & $m'm'm'$    & $mmm1'$    $(2221')$ \\
AFM9  & $P6_3'/m'm'c$ & $6'/m'm'm$  & $6'/m'm'm$ $(6'22')$ \\
AFM10 & $Pm'$         & $m'$        & $2'/m'$    $(2')$ \\
AFM11 & $P6_3/m'$     & $6/m'$      & $6/m1'$    $(61')$ \\
AFM12 & $P6_3/m'mc$   & $6/m'mm$    & $6/mmm1'$  $(6221')$ \\
AFM13 & $Cmc'm'$      & $mm'm'$     & $mm'm'$    $(2'2'2)$ \\
AFM14 & $Cm'cm'$      & $m'mm'$     & $m'mm'$    $(2'22')$ \\
AFM15 & $P6_3'/m'$    & $6'/m'$     & $6'/m'$    $(6')$ \\
AFM16 & $P6_3'/m'$    & $6'/m'$     & $6'/m'$    $(6')$ \\
AFM17 & $P6_3'/m'mc'$ & $6'/m'mm'$  & $6'/m'mm'$ $(6'22')$ \\
AFM18 & $Cm'cm'$      & $m'mm'$     & $m'mm'$    $(2'22')$ \\
AFM19 & $Cm'cm'$      & $m'mm'$     & $m'mm'$    $(2'22')$ \\
\else
\begin{tabular}{c|c|c}
label & magn. space group & magn. Laue group\\
\toprule
NM    & $P6_3/mmc$    & $6/mmm1'$  $(6221')$ \\
FM    & $P6_3/mm'c'$  & $6/mm'm'$  $(62'2')$ \\
AFM0  & $P6_3/m'm'c'$ & $6/mmm1'$  $(6221')$ \\
AFM1  & $Am'm'2$      & $m'm'm$    $(2'2'2)$ \\
AFM2  & $P\bar6'2c'$  & $6'/m'mm'$ $(6'22')$ \\
AFM3  & $Pm'$         & $2'/m'$    $ (2')$ \\
AFM4  & $Am'm'2$      & $m'm'm$    $(2'2'2)$ \\
AFM5  & $P6_3/m'$     & $6/m1'$    $(61')$ \\
AFM6  & $P6_3/m'$     & $6/m1'$    $(61')$ \\
AFM7  & $Am'a'2$      & $m'm'm$    $(2'2'2)$ \\
AFM8  & $Cm'c'm'$     & $mmm1'$    $(2221')$ \\
AFM9  & $P6_3'/m'm'c$ & $6'/m'm'm$ $(6'22')$ \\
AFM10 & $Pm'$         & $2'/m'$    $(2')$ \\
AFM11 & $P6_3/m'$     & $6/m1'$    $(61')$ \\
AFM12 & $P6_3/m'mc$   & $6/mmm1'$  $(6221')$ \\
AFM13 & $Cmc'm'$      & $mm'm'$    $(2'2'2)$ \\
AFM14 & $Cm'cm'$      & $m'mm'$    $(2'22')$ \\
AFM15 & $P6_3'/m'$    & $6'/m'$    $(6')$ \\
AFM16 & $P6_3'/m'$    & $6'/m'$    $(6')$ \\
AFM17 & $P6_3'/m'mc'$ & $6'/m'mm'$ $(6'22')$ \\
AFM18 & $Cm'cm'$      & $m'mm'$    $(2'22')$ \\
AFM19 & $Cm'cm'$      & $m'mm'$    $(2'22')$ \\
\fi
\end{tabular}
\caption{\label{tb:mlgsMn3Ge} Magnetic space and Laue groups of the magnetic
	structures shown in Fig.\,\ref{fig:Mn3GePhAFMx}, as well as of the non-magnetic 
	unit cell in Fig.~\ref{fig:Mn3Ge} (NM) and a ferromagnetic structure (FM) with 
	all moments along the [0001] direction. The magnetic Laue groups are
	given following the definition introduced by the present
	authors\cite{SKWE15a} as well as the one used by Kleiner\cite{Kle66}
	(in parentheses). The conventional setting concerning the sequence 
        of generators is used for the space groups and carried over to point and Laue
	groups.\cite{ITCA}}
\end{table}
%
The magnetic space \ifMPG, point,\fi and Laue groups corresponding to the structures
in Figs.~\ref{fig:Mn3Ge} and \ref{fig:Mn3GePhAFMx} as well as of a ferromagnetic alignment
of the Mn moments along the [0001] direction are given in Table~\ref{tb:mlgsMn3Ge}. For convenience
also the magnetic Laue group following the definition of Kleiner\cite{Kle66} is 
given in parentheses. As can be seen, a number of structures have the same magnetic
space group and moreover a number of these lead to the same magnetic Laue groups. 

The corresponding symmetry-restricted tensor shapes are:\cite{Kle66,SKWE15a}\\
%
\begin{equation}
  \label{eq:sigma-Mn3Ge_PhNM}
  \underline{\sigma}^{\rm NM} =
  \left(
    \begin{array}{ccc}
      \sigma_{xx} & 0           & 0 \\
      0           & \sigma_{xx} & 0 \\
      0           & 0           & \sigma_{zz}
    \end{array}
  \right)
\end{equation}
%
%
\begin{equation}
  \label{eq:sigma-Mn3Ge_PhFM}
  \underline{\sigma}^{\rm FM} =
  \left(
    \begin{array}{ccc}
      \sigma_{xx} & \sigma_{xy} & 0 \\
     -\sigma_{xy}  & \sigma_{xx} & 0 \\
      0           & 0           & \sigma_{zz}
    \end{array}
  \right)
\end{equation}
%
%
\begin{equation}
  \label{eq:sigma-Mn3Ge_PhAFM0}
  \underline{\sigma}^{\rm AFM0,2,5,6,9,11,12,15,16,17} =
  \left(
    \begin{array}{ccc}
      \sigma_{xx} & 0           & 0 \\
      0           & \sigma_{xx} & 0 \\
      0           & 0           & \sigma_{zz}
    \end{array}
  \right)
\end{equation}
%
%
\begin{equation}
  \label{eq:sigma-Mn3Ge_PhAFM1}
  \underline{\sigma}^{\rm AFM1,4,13,18,19} =
  \left(
    \begin{array}{ccc}
      \sigma_{xx} & 0           & \sigma_{xz} \\
      0           & \sigma_{yy} & 0 \\
     -\sigma_{xz} & 0           & \sigma_{zz}
    \end{array}
  \right)
\end{equation}
%
%
\begin{equation}
  \label{eq:sigma-Mn3Ge_PhAFM3}
  \underline{\sigma}^{\rm AFM3,10} =
  \left(
    \begin{array}{ccc}
      \sigma_{xx} &  \sigma_{xy} & \sigma_{xz} \\
      \sigma_{xy} &  \sigma_{yy} & \sigma_{yz} \\
     -\sigma_{xz} & -\sigma_{yz} & \sigma_{zz}
    \end{array}
  \right)
\end{equation}
%
%
\begin{equation}
  \label{eq:sigma-Mn3Ge_PhAFM7}
  \underline{\sigma}^{\rm AFM7,AFM14} =
  \left(
    \begin{array}{ccc}
      \sigma_{xx} & 0           & 0 \\
      0           & \sigma_{yy} & \sigma_{xz} \\
      0           &-\sigma_{xz} & \sigma_{zz}
    \end{array}
  \right)
\end{equation}
%
%
\begin{equation}
  \label{eq:sigma-Mn3Ge_PhAFM8}
  \underline{\sigma}^{\rm AFM8} =
  \left(
    \begin{array}{ccc}
      \sigma_{xx} & 0           & 0 \\
      0           & \sigma_{yy} & 0 \\
      0           & 0           & \sigma_{zz}
    \end{array}
  \right) \; \mathrm{.}
\end{equation}
%
In fact, for the twenty actual and hypothetical structures in 
Fig.~\ref{fig:Mn3GePhAFMx} obviously only four different tensor shapes are 
found: a) The non-magnetic shape of Eq.~\eqref{eq:sigma-Mn3Ge_PhNM} with the 
crystallographic anisotropy $\sigma_{xx} = \sigma_{yy} \neq \sigma_{zz}$ that  
applies for a number of antiferromagnetic structures [see 
Eq.~\eqref{eq:sigma-Mn3Ge_PhAFM0}], b) the ferromagnetic shape of 
Eq.~\eqref{eq:sigma-Mn3Ge_PhFM} with an additional off-diagonal anti-symmetric 
element representing the AHE. This is found again for five 
antiferromagnetic structures in Eq.~\eqref{eq:sigma-Mn3Ge_PhAFM1} and two others 
in Eq.~\eqref{eq:sigma-Mn3Ge_PhAFM7}, only with different principal axes, 
c) the full tensor with two anomalous 
Hall conductivities in Eq.~\eqref{eq:sigma-Mn3Ge_PhAFM3}, and finally d) the 
diagonal fully anisotropic form of AFM8 in Eq.~\eqref{eq:sigma-Mn3Ge_PhAFM8}.
Note that all configurations assumed in experimental (AFM13\cite{NKH15,KTN16}, 
AFM14\cite{KTN16,NFS+16}, AFM18/19\cite{NFS+16}) as well as theoretical 
(AFM1-4\cite{KF14a}, AFM18/19\cite{ZZFY17a,YSZ+17,ZZS+17,ZSY+17}) investigations
on the anomalous Hall effect in hexagonal Mn$_3X$ compounds are found to exhibit
at least one off-diagonal anti-symmetric element of $\underline{\mbox{\boldmath$\sigma$}}$, 
with the exception of AFM1 for which no AHE is predicted in Ref.~\onlinecite{KF14a}.

Corresponding tensor forms for direct and inverse spin conductivity can be
found in Ref.~\onlinecite{SKWE15a}, those for the spin-orbit torkance and the
Edelstein polarization in chiral structures are given in
Refs.~\onlinecite{WCS+16a} and \onlinecite{WCE18}, respectively.




\subsection{X-ray absorption spectroscopy}

\begin{figure}[p]
 \begin{center}
 \includegraphics[angle=0,width=0.7\linewidth,clip]{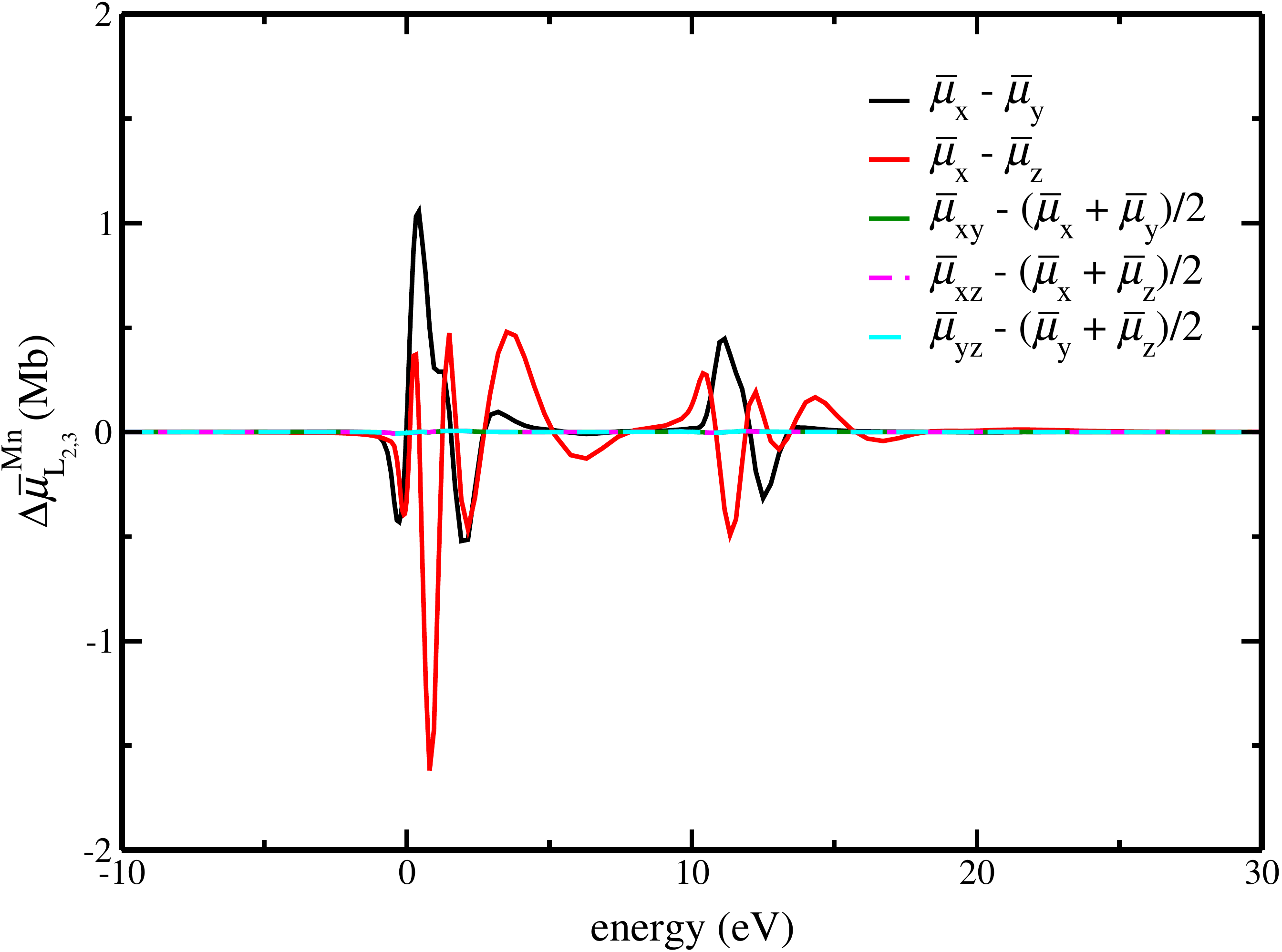}
 \includegraphics[angle=0,width=0.7\linewidth,clip]{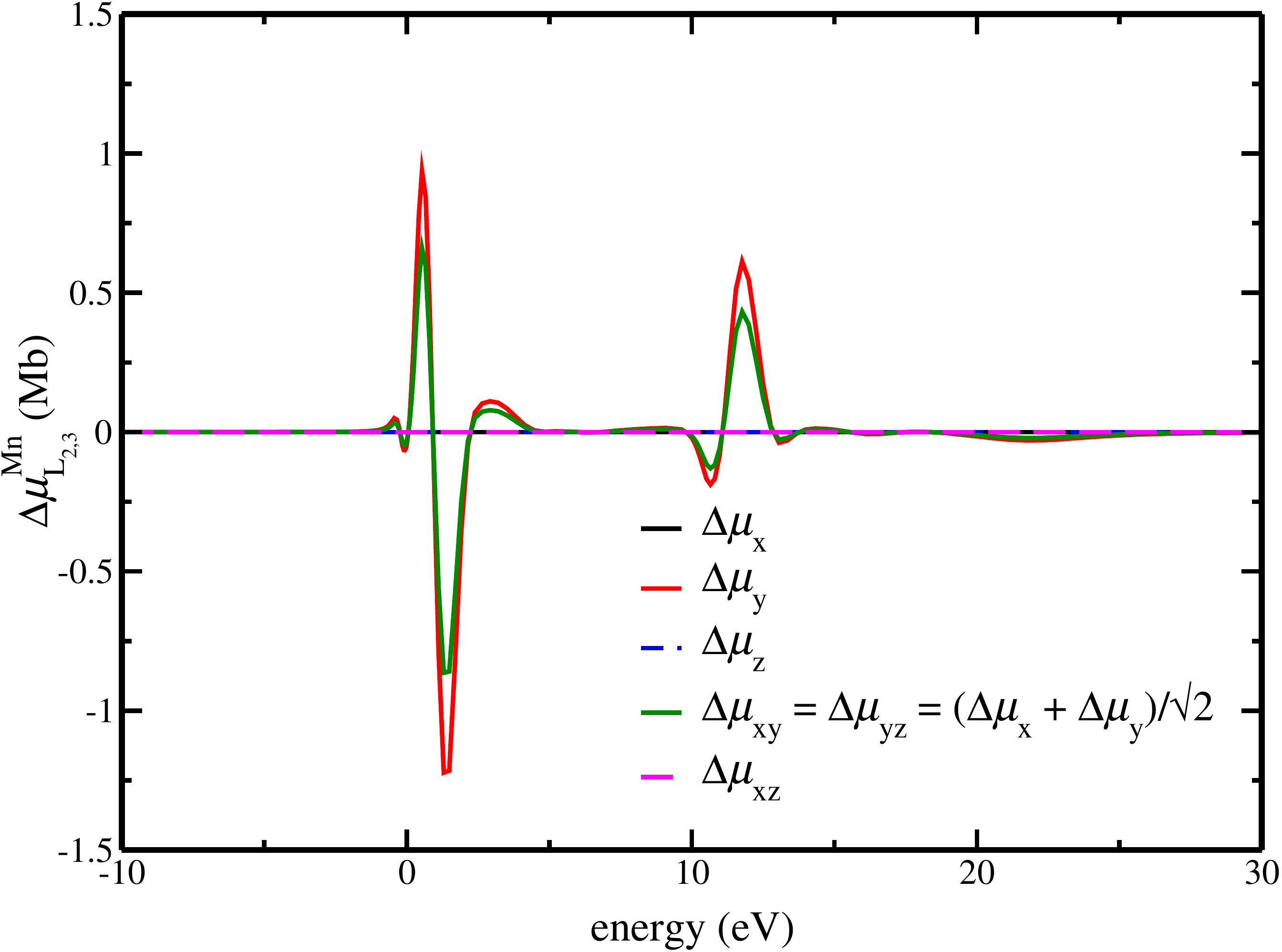}
  \caption{\label{FIG-XAS-Mn3GePhAFM1} Top: Differential polarization-averaged 
X-ray absorption spectra $\Delta\bar\mu$ comparing incidence along high-symmetry directions in 
Mn$_3$Ge with AFM1 magnetic structure (see Fig.~\ref{fig:Mn3GePhAFMx}). Bottom: 
Corresponding X-ray magnetic circular dichroism (XMCD) spectra $\Delta\mu$.}
\end{center}
\end{figure}

\begin{figure}[p]
 \begin{center}
 \includegraphics[angle=0,width=0.7\linewidth,clip]{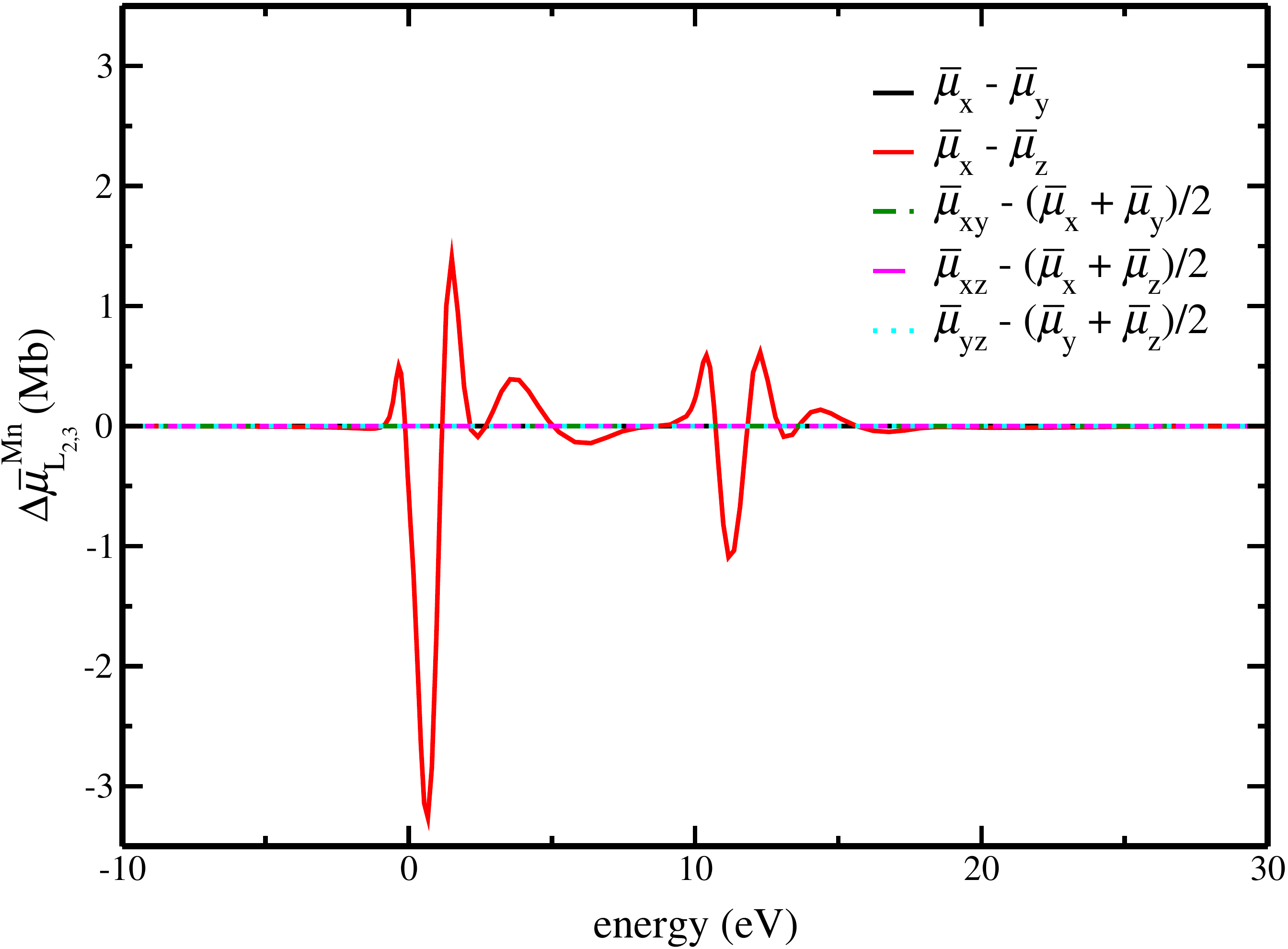}
 \includegraphics[angle=0,width=0.7\linewidth,clip]{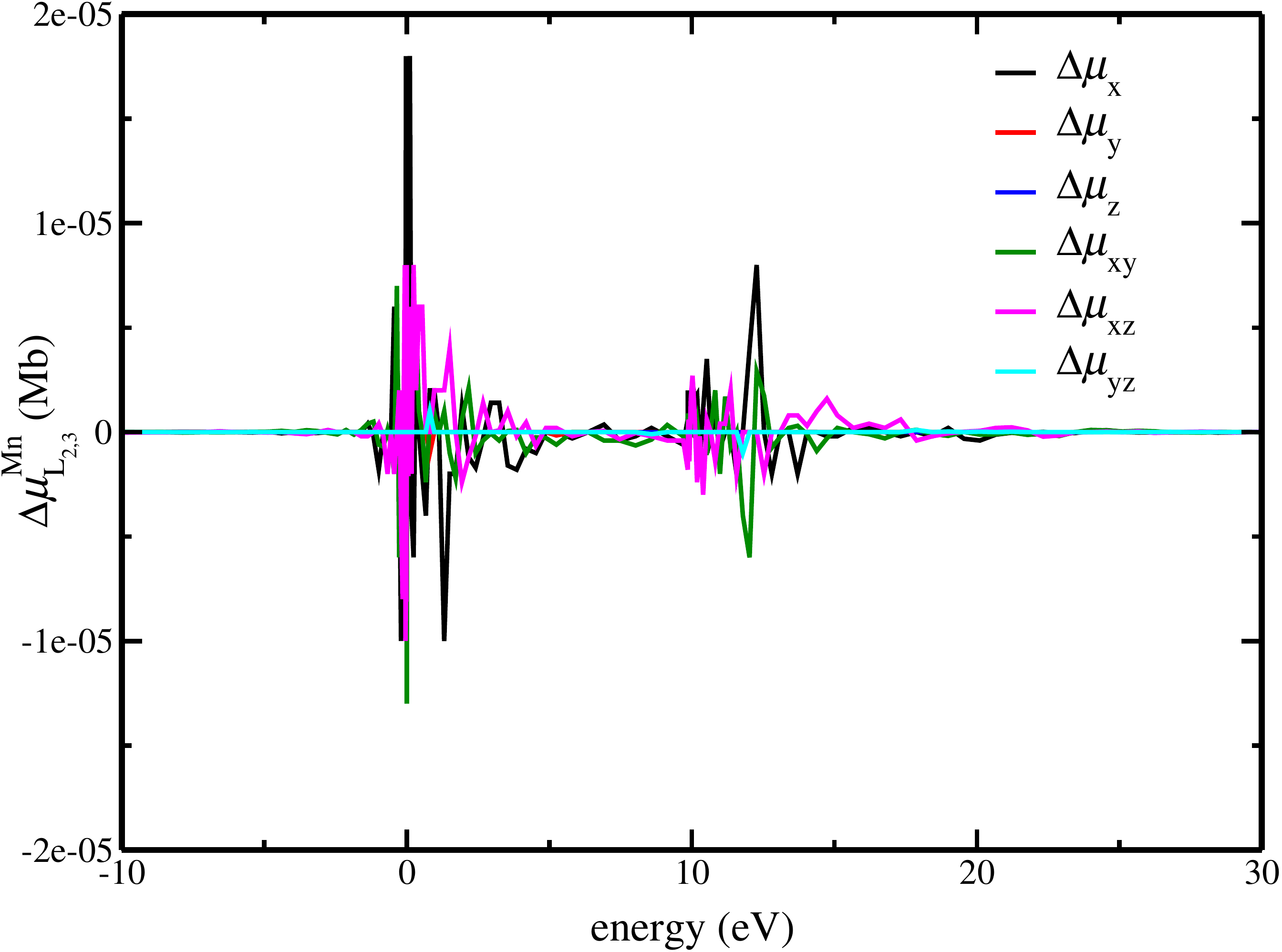}
  \caption{\label{FIG-XAS-Mn3GePhAFM2} Top: Differential polarization-averaged 
X-ray absorption spectra $\Delta\bar\mu$ comparing incidence along high-symmetry directions in 
Mn$_3$Ge with AFM2 magnetic structure (see Fig.~\ref{fig:Mn3GePhAFMx}). Bottom: 
Corresponding X-ray magnetic circular dichroism (XMCD) spectra $\Delta\mu$.}
\end{center}
\end{figure}

\begin{figure}[p]
 \begin{center}
 \includegraphics[angle=0,width=0.7\linewidth,clip]{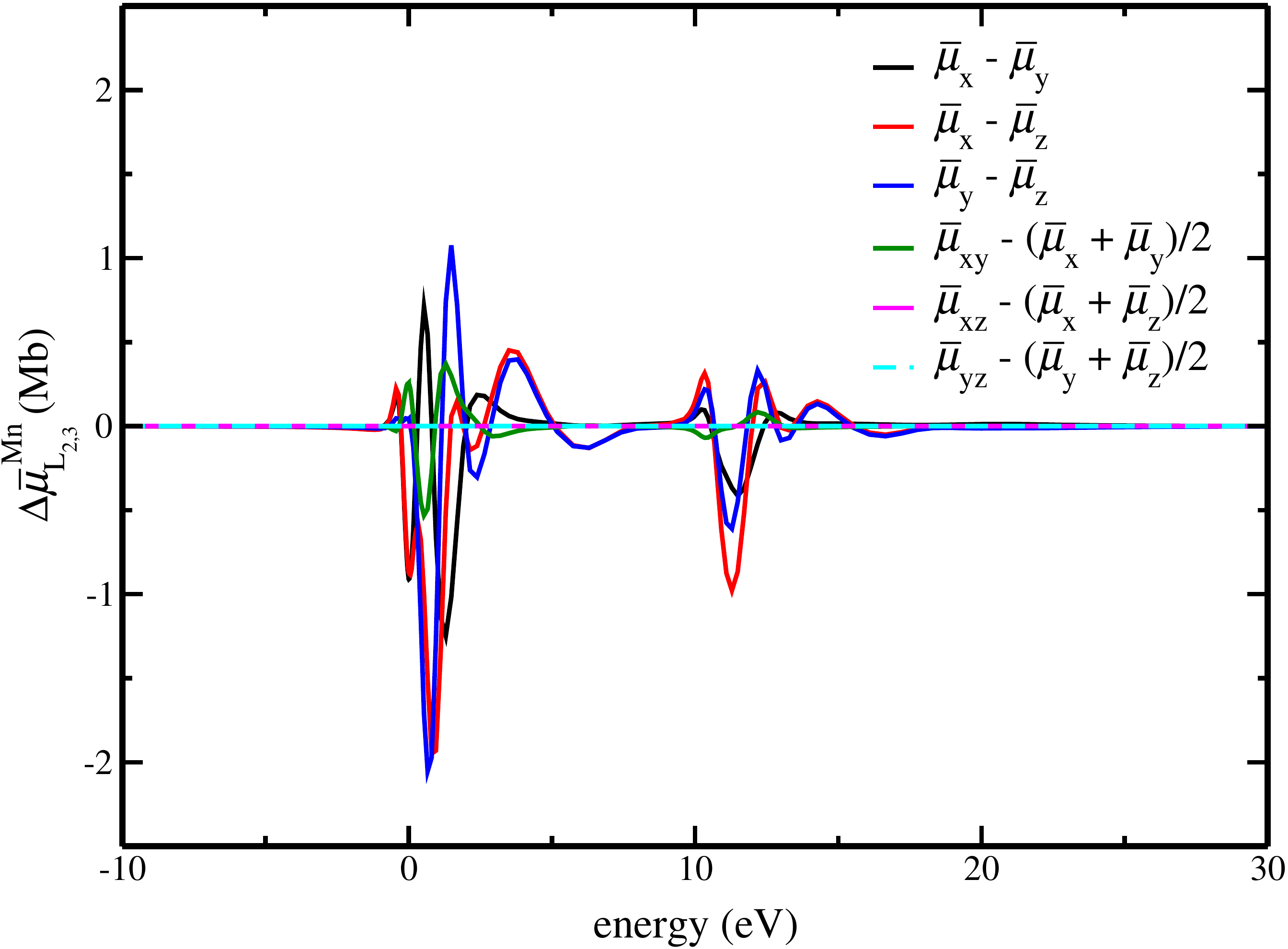}
 \includegraphics[angle=0,width=0.7\linewidth,clip]{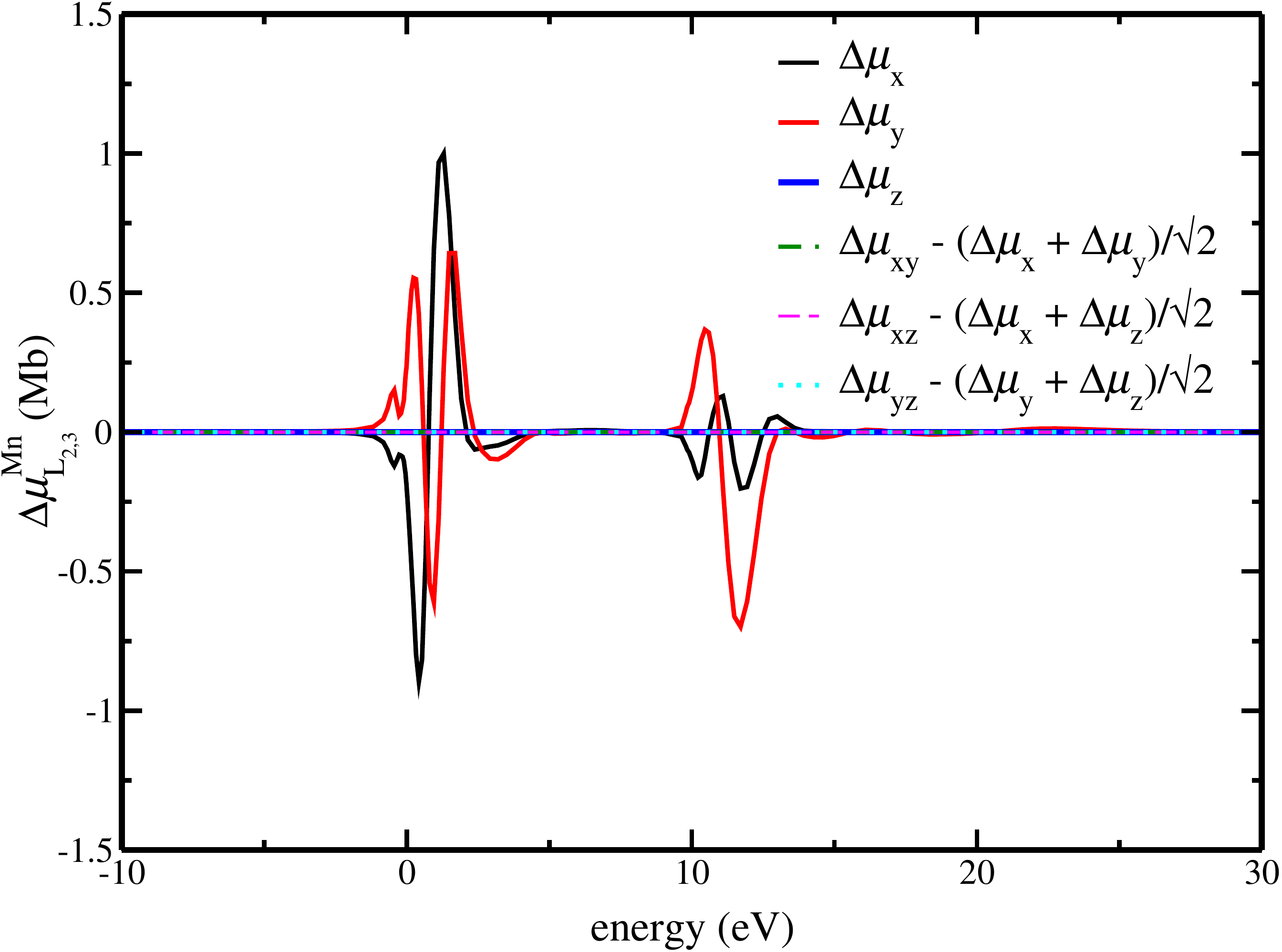}
  \caption{\label{FIG-XAS-Mn3GePhAFM3} Top: Differential polarization-averaged 
X-ray absorption spectra $\Delta\bar\mu$ comparing incidence along high-symmetry directions in 
Mn$_3$Ge with AFM3 magnetic structure (see Fig.~\ref{fig:Mn3GePhAFMx}). Bottom: 
Corresponding X-ray magnetic circular dichroism (XMCD) spectra $\Delta\mu$.}
\end{center}
\end{figure}

\begin{figure}[p]
 \begin{center}
 \includegraphics[angle=0,width=0.7\linewidth,clip]{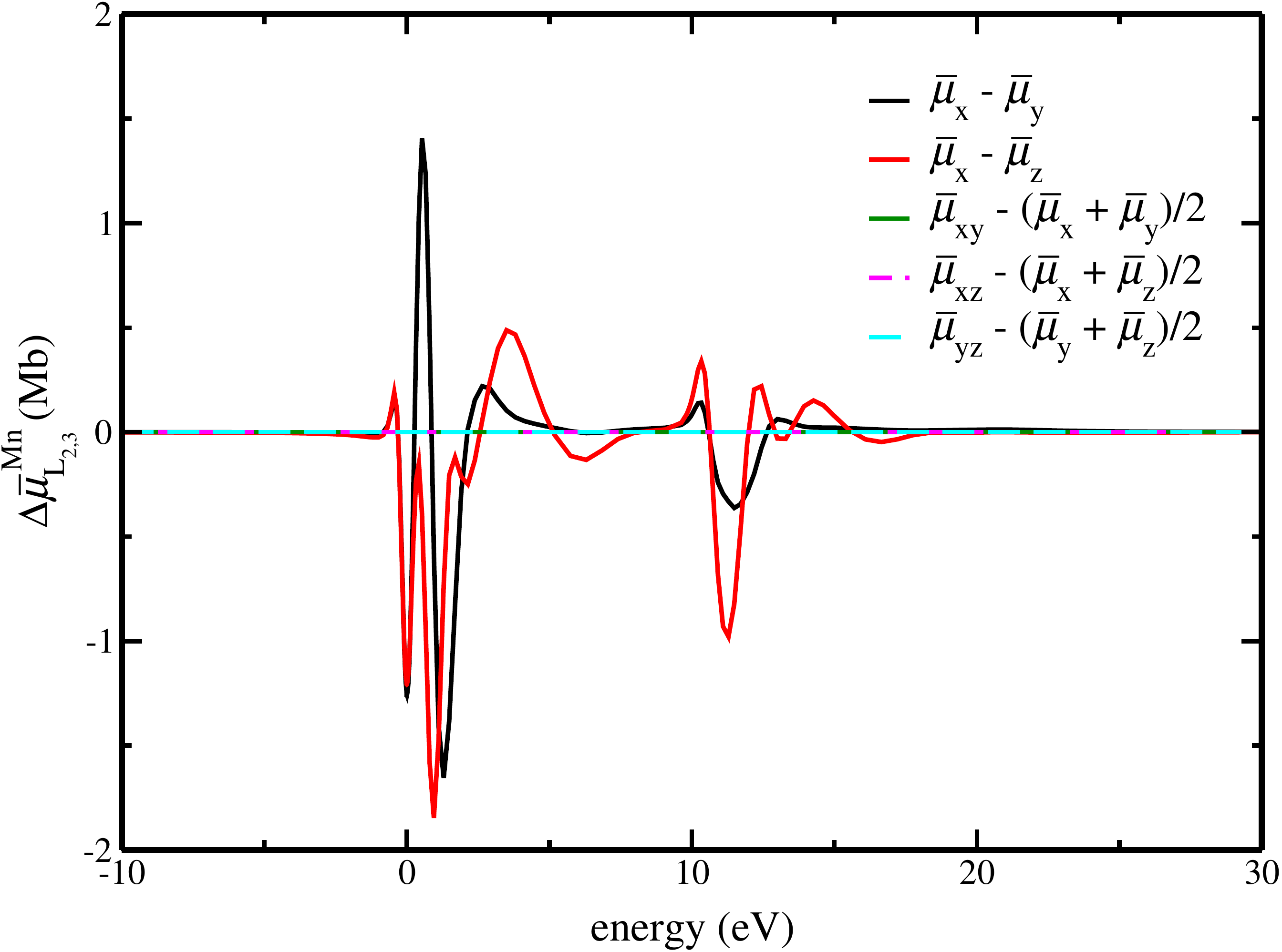}
 \includegraphics[angle=0,width=0.7\linewidth,clip]{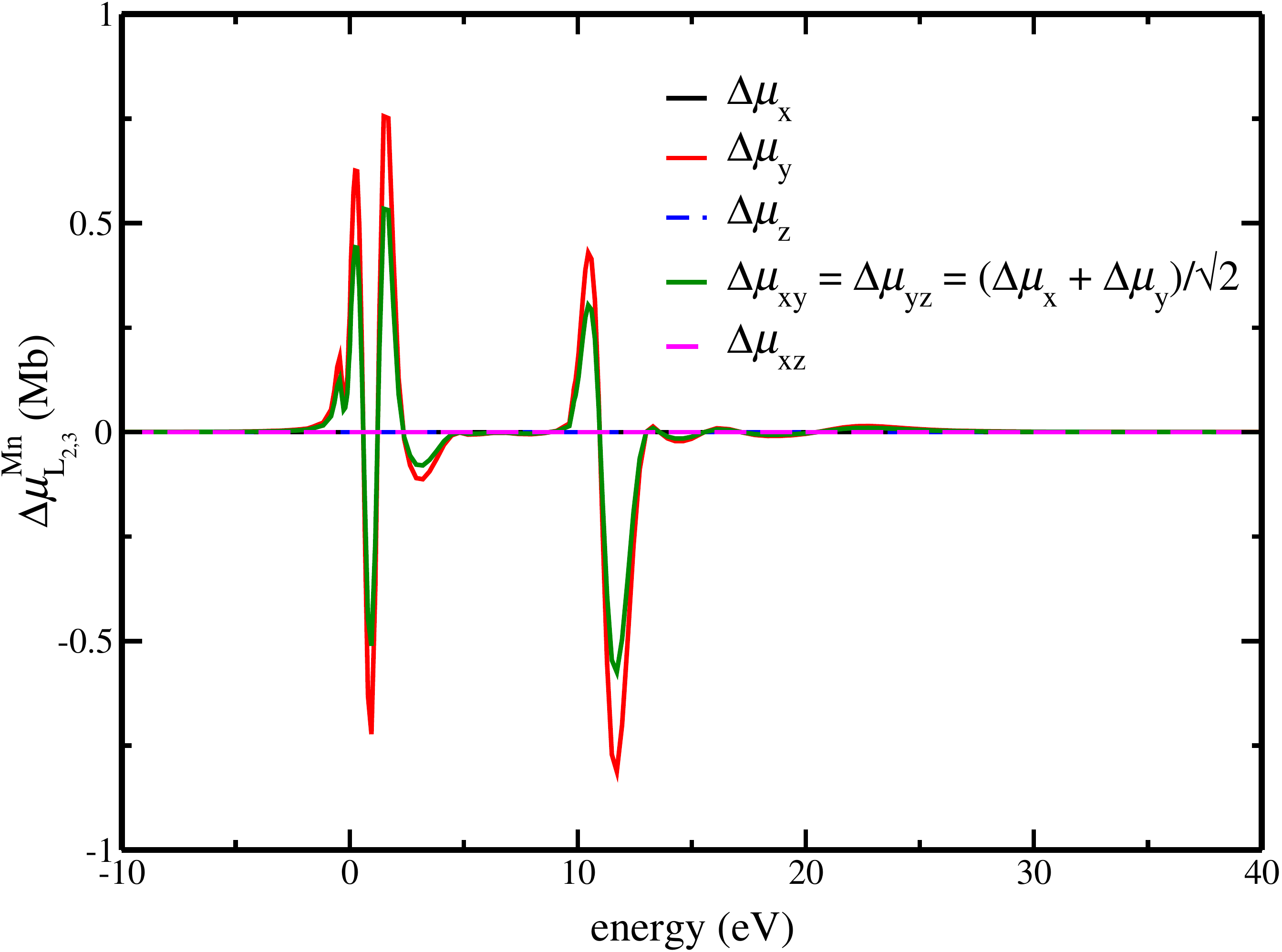}
  \caption{\label{FIG-XAS-Mn3GePhAFM4} Top: Differential polarization-averaged 
X-ray absorption spectra $\Delta\bar\mu$ comparing incidence along high-symmetry directions in 
Mn$_3$Ge with AFM4 magnetic structure (see Fig.~\ref{fig:Mn3GePhAFMx}). Bottom: 
Corresponding X-ray magnetic circular dichroism (XMCD) spectra $\Delta\mu$.}
\end{center}
\end{figure}

The tensor shapes presented above were confirmed for selected representative 
cases using first-principles calculations of X-ray absorption spectra for 
circularly polarized light. 
Figures~\ref{FIG-XAS-Mn3GePhAFM1}-\ref{FIG-XAS-Mn3GePhAFM4} show for the 
structures AFM1-4 the difference in polarization-averaged spectra, 
$\Delta\bar\mu$, along high-symmetry (Cartesian and intermediate) directions in 
the top panels and the corresponding spectra giving the difference in absorption 
for left and right circularly polarized X-rays, the XMCD signals, in the bottom 
panels. As stated above (see section~\ref{ssec:XAS-Mn3Ir}), a finite 
$\Delta\bar\mu$ signal in the upper panels of 
Figs.~\ref{FIG-XAS-Mn3GePhAFM1}-\ref{FIG-XAS-Mn3GePhAFM4} indicates an 
anisotropy in the symmetric part of the optical conductivity tensor, i.e., a 
difference on the diagonal and/or presence of symmetric contributions to 
off-diagonal elements. For the XMCD spectra in the lower panel a finite signal 
$\Delta\mu$ along a Cartesian direction $i$ confirms the presence of 
anti-symmetric off-diagonal elements of 
$\underline{\mbox{\boldmath$\sigma$}}(\omega)$ with indices $j,k \neq i$. 
Moreover, an XMCD for incidence along intermediate directions $\langle 110 
\rangle$ is in all cases, if present, found to be a linear combination of the 
corresponding signals for incidence along the Cartesian axes. Taken together, in 
all four cases the predicted tensor shapes are confirmed.
%
%
%
%
%
%
%
%


\section{Conclusions}

Based on general and model-independent symmetry arguments for the occurrence of
galvano-magnetic and magneto-optical phenomena in materials with arbitrary 
magnetic order, the implications of in particular off-diagonal anti-symmetric 
elements of the frequency-dependent conductivity tensor in the coplanar 
non-collinear antiferromagnets Mn$_3X$ with $X = $ Ir and Ge have been investigated.
For the cubic Mn$_3$Ir-type compound the results of first-principles Kubo linear
response calculations of the anomalous and spin Hall conductivities could be shown
to be in reasonable agreement with previous findings. In addition the concentration
dependence of the anomalous and spin Hall effects in substitutionally disordered
alloys is found to be non-trivial in particular concerning the AHE. A clear increase of
the intrinsic values for both, AHE and SHE, with increasing atomic number in the
pure compounds Mn$_3$Rh, Mn$_3$Ir, and Mn$_3$Pt is however in accordance with 
expectations concerning the relevance of spin-orbit coupling. The magneto-optical
properties of pure Mn$_3$Ir are again found to be in agreement with the shape
of the optical conductivity tensor as well as with previous theoretical work. 
Comparison with corresponding results for the diagonal and off-diagonal
optical conductivities in bcc Fe moreover shows that the magneto-optical
Kerr effect is of comparable magnitude in both. Calculated polarization-averaged X-ray 
absorption and magnetic circular dichroism spectra confirm expectations concerning
their symmetry and in addition the size of the XMCD signal suggests the possibility
of experimental confirmation. For the hexagonal Heusler compound Mn$_3$Ge a number
of proposed as well as additional hypothetical spin-compensated non-collinear 
configurations has been studied w.r.t.\ to magnetic symmetry and consequent 
electrical conductivity tensor shape. The occurrence of at least one independent
anomalous Hall conductivity can be confirmed for all magnetic structures usually 
assumed in the experimental and theoretical literature. For selected representative
cases the tensor shapes have been confirmed by supplementary X-ray absorption spectroscopy
calculations. The implications of a non-coplanar spin texture in chiral and 
achiral magnetic structures of Mn$_3$Ge on the linear response properties will be
discussed in a separate contribution.\cite{WME18}




\begin{acknowledgments}
	Financial support by the DFG via SFB~1277 (\emph{Emer-
	gente relativistische Effekte in der Kondensierten Ma-
	terie}) is gratefully acknowledged. 
	This work was further supported by the
	European Regional Development Fund (ERDF), project
	CEDAMNF, reg.\ no.\ CZ.02.1.01/0.0/0.0/15\_003/0000358.
\end{acknowledgments}



\begin{thebibliography}{63}
\bibitem [{VES()}]{VESTA}%
  \BibitemOpen
  \href@noop {} {}\bibinfo {howpublished} {This figure has been created using
  the software VESTA\cite{MI11}.}\BibitemShut {Stop}%
\end{thebibliography}

%


\end{document}